\documentclass[twocolumn,twocolappendix,trackchanges]{aastex631}

\usepackage{physics}
\usepackage{amsmath}
\usepackage{tabu}
\usepackage{booktabs}
\usepackage{subfigure}
\usepackage{hyperref}
\usepackage{ulem}
\usepackage{enumerate}
\usepackage{CJK}

\newcommand{\sect}{Section~}

\newcommand{\figu}{Figure~}
\newcommand{\figus}{Figures~}

\newcommand{\eq}{Equation~}

\newcommand{\athenak}{\texttt{AthenaK}}
\newcommand{\athena}{\texttt{Athena++}}
\newcommand{\kokkos}{\texttt{Kokkos}}

\shorttitle{MHD Accretion onto BH in Ellipticals}
\shortauthors{Guo et al.}
\graphicspath{{./}{figures/}}
\begin{document}

\title{Magnetized Accretion onto and Feedback from Supermassive Black Holes in Elliptical Galaxies}

\begin{CJK*}{UTF8}{gbsn}

\correspondingauthor{Minghao Guo}
\email{mhguo@princeton.edu}

\author[0000-0002-3680-5420]{Minghao Guo (郭明浩)}
\affiliation{Department of Astrophysical Sciences, Princeton University, Princeton, NJ 08544, USA}

\author[0000-0001-5603-1832]{James M. Stone}
\affiliation{School of Natural Sciences, Institute for Advanced Study, 1 Einstein Drive, Princeton, NJ 08540, USA}
\affiliation{Department of Astrophysical Sciences, Princeton University, Princeton, NJ 08544, USA}

\author[0000-0001-9185-5044]{Eliot Quataert}
\affiliation{Department of Astrophysical Sciences, Princeton University, Princeton, NJ 08544, USA}

\author[0000-0003-2896-3725]{Chang-Goo Kim}
\affiliation{Department of Astrophysical Sciences, Princeton University, Princeton, NJ 08544, USA}

\begin{abstract}
We present three-dimensional magnetohydrodynamic (MHD) simulations of the fueling of supermassive black holes in elliptical galaxies from a turbulent cooling medium on galactic scales, taking M87* as a typical case. We find that the mass accretion rate is increased by a factor of $\sim 10$ compared with analogous hydrodynamic simulations. The scaling of $\dot{M} \sim r^{1/2}$ roughly holds from $\sim 10\,\mathrm{pc}$ to $\sim 10^{-3}\,\mathrm{pc}$ ($\sim 10\, r_\mathrm{g}$) with the accretion rate through the event horizon being $\sim 10^{-2}\, M_\odot\,\mathrm{yr^{-1}}$. The accretion flow on scales $\sim 0.03-3\,\mathrm{kpc}$ takes the form of magnetized filaments. Within $\sim 30\,\mathrm{pc}$, the cold gas circularizes, forming a highly magnetized ($\beta\sim 10^{-3}$) thick disk supported by a primarily toroidal magnetic field. The cold disk is truncated and transitions to a turbulent hot accretion flow at $\sim0.3\,\mathrm{pc}$ ($10^3\,r_\mathrm{g}$). There are strong outflows towards the poles driven by the magnetic field. The outflow energy flux increases with smaller accretor size, reaching $\sim 3\times10^{43}\,\mathrm{erg\,s^{-1}}$ for $r_\mathrm{in}=8\,r_\mathrm{g}$; this corresponds to a nearly constant energy feedback efficiency of $\eta\sim0.05-0.1$ independent of accretor size. The feedback energy is enough to balance the total cooling of the M87/Virgo hot halo out to $\sim 50$ kpc. The accreted magnetic flux at small radii is similar to that in magnetically arrested disk models, consistent with the formation of a powerful jet on horizon scales in M87. Our results motivate a subgrid model for accretion in lower-resolution simulations in which the hot gas accretion rate is suppressed relative to the Bondi rate by $\sim (10r_\mathrm{g}/r_\mathrm{B})^{1/2}$.
\end{abstract}


\section{Introduction} \label{sec:intro}

Supermassive black holes (SMBHs), harbored in the nuclei of almost all massive galaxies, correlate with properties of their hosts~\citep{Magorrian1998AJ....115.2285M, Ferrarese2000ApJ...539L...9F, Gebhardt2000ApJ...539L..13G, Kormendy&Ho2013ARA&A..51..511K}. How these black holes accrete gas, grow, and feed back mass, momentum, and energy into their environments remain crucial unsolved problems. Accurately modeling feeding and feedback from galactic to event horizon scales is a formidable task; spatial scales spanning nearly nine orders of magnitude (from mpc to Mpc) need to be resolved~\citep{Gaspari2020NatAs...4...10G}. 

The exact process by which SMBHs at the centers of galaxies receive their fuel is not yet fully understood. In massive galaxies, the gravitational radius of a black hole is typically $\sim 10^6$ times smaller than the radius where adiabatic accretion begins, known as the Bondi radius~\citep{Bondi1952MNRAS.112..195B}. The Bondi flow at larger radii gives way to a hot radiatively inefficient accretion flow (RIAF) at smaller radii when the cooling time scale is longer than the accretion time scale~\citep{Ichimaru1977ApJ...214..840I, Narayan&Yi1995ApJ...452..710N, Yuan2014ARA&A..52..529Y}. However, under some circumstances, gas at larger radii can cool, leading to the formation of a thin disk or chaotic cold accretion flow that significantly changes the accretion flow and increases the accretion rate~\citep{Li&Bryan2012ApJ...747...26L, Sharma2012, Gaspari2013MNRAS.432.3401G}. These multiphase flows are observed in the cores of many galaxy clusters and massive galaxies~\citep[e.g.,][]{McDonald2012ApJ...746..153M, Tremblay2016Natur.534..218T, Combes2019A&A...623A..79C, Boselli2019A&A...623A..52B, Li2020ApJ...889L...1L}.

Contemporary event horizon-scale general relativistic magnetohydrodynamic simulations (GRMHD)~\citep{Gammie2003ApJ...589..444G, Narayan2012MNRAS.426.3241N, Porth2019ApJS..243...26P, White2020ApJ...891...63W} set idealized initial conditions, e.g., a torus in hydrodynamic equilibrium~\citep{Fishbone&Moncrief1976ApJ...207..962F, Kozlowski1978A&A....63..209K}. More realistic initial and boundary conditions on galactic scales may help to construct a more consistent model of black hole accretion at smaller radii. Connecting the large and small scales in this way is also critical for developing more physical models of black hole growth and feedback in cosmological simulations that lack the physics or resolution to follow the gas to small radii~\citep{Hopkins&Quataert2010MNRAS.407.1529H, Hopkins&Quataert2011MNRAS.415.1027H, Li&Bryan2014ApJ...789..153L}. Recent years have seen considerable attempts to link these scales in various environments with different techniques, including ``super-Lagrangian'' or ``hyper-refinement'' methods~\citep[e.g.,][]{Angles2021ApJ...917...53A, Hopkins2024OJAp....7E..18H, Hopkins2024OJAp....7E..19H}, ``zoom-in'' using nested meshes~\citep[e.g.,][]{Ressler2018MNRAS.478.3544R, Ressler2020ApJ...896L...6R}, direct simulations assuming smaller scale separation~\citep[e.g.,][]{Lalakos2022ApJ...936L...5L, Kaaz2023ApJ...950...31K, Olivares2023A&A...678A.141O, Lalakos2024ApJ...964...79L}, or a ``multi-zone'' method that attempts to pass information both from large to small scales and vice-versa~\citep{Cho2023ApJ...959L..22C, Cho2024arXiv240513887C}. 

In~\citet{Guo2023ApJ...946...26G}, we performed a suite of nested-mesh hydrodynamic simulations with radiative cooling and heating to resolve the multi-scale multi-phase accretion flow and bridge the gap between galactic scales and the event horizon. In those runs, accretion takes the form of multiphase gas at radii less than about a kpc. Cold gas accretion includes two dynamically distinct stages: a disk stage in which the cold gas mostly resides in a rotationally supported disk, and relatively rare chaotic stages in which the cold gas inflows via chaotic streams. The accretion rate scales with radius as $\dot{M}\propto r^{1/2}$ when hot gas dominates. For the specific case of M87, we obtain an accretion rate similar to what is inferred from the Event Horizon Telescope observations~\citep{M87EHT_I_2019ApJ...875L...1E}. A significant limitation of the hydrodynamic simulations is that they do not include the dynamical effects of magnetic fields, which are especially important for the transport of angular momentum via, e.g., magnetorotational instability~\citep[MRI;][]{Balbus1991ApJ...376..214B, Balbus1998RvMP...70....1B}, launching of winds and outflows~\citep{Blandford1982MNRAS.199..883B}, and forming structures such as a magnetically arrested disk (MAD)~\citep{Narayan2003PASJ...55L..69N}. Here we extend our previous work by performing a new suite of multi-scale magnetohydrodynamic (MHD) models of black hole accretion in elliptical galaxies. The simulations focus on understanding the magnetized accretion and outflows in such models; an important limitation is that we cannot run long enough to self-consistently study the effects of feedback on the larger scale solution (see \citealt{Cho2023ApJ...959L..22C, Cho2024arXiv240513887C} for an attempt to do so in GRMHD simulations of adiabatic Bondi-like accretion).

The rest of this article is organized as follows. In \sect\ref{sec:method} we describe the numerical model we adopt. \sect\ref{sec:results} presents the results of the MHD simulations. In \sect\ref{sec:discussion}, we discuss the implications of our results. We conclude in \sect\ref{sec:summary}. 

\section{Method} \label{sec:method}

We perform magnetohydrodynamic simulations using \athenak~(J. M. Stone et al. 2024, in preparation), a performance portable version of the \athena~\citep{Stone2020ApJS..249....4S} code implemented using the \kokkos\ library~\citep{Trott2021CSE....23e..10T}. \athenak\ provides a variety of reconstruction methods, Riemann solvers, and integrators for solving the MHD equations. In our simulations, we adopt the piecewise linear (PLM) reconstruction method, the HLLD Riemann solver, and the RK2 time integrator to solve the MHD equations. The adaptive mesh refinement (AMR) in \athenak\ allows us to flexibly achieve a high resolution and good performance over an extremely large dynamic range. The setup is an extension of the purely hydrodynamic model in~\citet{Guo2023ApJ...946...26G}, where we presented the method in detail.

In brief, the equations solved in our simulations are
\begin{align}
    \frac{\partial \rho}{\partial t}+\nabla \cdot(\rho \boldsymbol{v}) &=0, \label{eq:mhd_mass_eq}\\
    \frac{\partial \rho \boldsymbol{v}}{\partial t}+\nabla \cdot\left(P_\mathrm{tot}\mathbf{I}+\rho \boldsymbol{v} \boldsymbol{v}-\boldsymbol{B} \boldsymbol{B}\right) &=\rho\boldsymbol{g}, \label{eq:mhd_momentum_eq}\\
    \frac{\partial E}{\partial t}+\nabla \cdot\left[\left(E+P_\mathrm{tot}\right) \boldsymbol{v}-\boldsymbol{B}\boldsymbol{B} \cdot \boldsymbol{v}\right] &=\rho\boldsymbol{g}\cdot\boldsymbol{v}-q_-+q_+, \label{eq:mhd_energy_eq}\\
    \frac{\partial \boldsymbol{B}}{\partial t}-\nabla \times(\boldsymbol{v} \times \boldsymbol{B}) &=0, \label{eq:mhd_induction_eq}
\end{align}
where $\rho$ is the gas density, $\boldsymbol{v}$ is the velocity, $\boldsymbol{B}$ is the magnetic field, $P_\mathrm{tot}=P_\mathrm{gas}+B^2/2$ is the total pressure including both thermal and magnetic contribution, $\boldsymbol{g}$ is the gravitational acceleration, $E=E_\text{int}+\rho v^2/2+B^2/2$ is the total energy density with $E_\text{int}=P_\mathrm{gas}/(\gamma-1)$ the internal energy density, $q_-$ is cooling rate per unit volume caused by optically thin bremsstrahlung and line cooling for solar metallicity, and $q_+$ is ad hoc heating rate compensating the cooling shell by shell to keep a global equilibrium but allow local thermal instability. These equations are written in units such that the magnetic permeability $\mu_\mathrm{m}=1$. The initial and boundary conditions, gravitational field, cooling, heating, floors, and flux corrections are similar to \citet{Guo2023ApJ...946...26G}. We briefly present the model and major extension below.

The initial conditions of gas number density, $n_\mathrm{init}(r)$, and temperature, $T_\mathrm{init}(r)$, are obtained by solving 1D hydrostatic equilibrium assuming a flat cored entropy profile following~\citet{Martizzi2019MNRAS.483.2465M}. To better match the observations~\citep{Urban2011MNRAS.414.2101U, Russell2015MNRAS.451..588R}, we adopt a set of parameters slightly different from the previous hydrodynamic case. We use the radius of entropy core $r_0=1\,\mathrm{kpc}$, the normalization of entropy $K_0=3.3\,\mathrm{keV\,cm^{2}}$, the large-scale slope of the entropy profile $\xi=1.1$, and number density $n_\mathrm{init}(r=r_0)=0.5\,\mathrm{cm^{-3}}$. The gravitational potential includes contribution from the SMBH of mass $M_\mathrm{BH}=6.5\times10^9\, M_\odot$~\citep{M87EHT_I_2019ApJ...875L...1E} corresponding to a gravitational radius $r_\mathrm{g}\equiv GM_\mathrm{BH}/c^2\approx0.31\,\mathrm{mpc}$, star by a NFW profile~\citep{Navarro1997ApJ...490..493N} with the stellar mass $M_\mathrm{S}=3\times10^{11}\,M_\odot$ and characteristic radius $r_\mathrm{S}=2\mathrm{kpc}$, and dark matter by another NFW profile with mass $M_\mathrm{DM}=10^{14}\,M_\odot$ and radius $r_\mathrm{DM}=110\,\mathrm{kpc}$. Note that $M_\mathrm{DM}$ is not the ``virial mass'' or ``cluster mass'' of the cluster since it is only the mass inside some radius that is actually smaller compared to the virial radius. This setup gives a Bondi radius $r_\mathrm{B}=2GM_\mathrm{BH}/c_\mathrm{s}^2\approx120\,\mathrm{pc}$ and a Bondi accretion rate $\dot{M}_\mathrm{B}\approx 0.15\, M_\odot\,\mathrm{yr^{-1}}$. The simulations are initialized with random isobaric density perturbations and Gaussian random vector potential field $\boldsymbol{A}$ with $\boldsymbol{B}=\nabla\times\boldsymbol{A}$ to seed turbulence and model entangled magnetic fields on large scales. In practice, we set the perturbation and magnetic field only on large scales and let it naturally cascade to smaller scales during the evolution. In the fiducial suite of runs, we set $\delta\rho/\rho\sim0.1$, plasma $\beta\approx100$, and wavelength $10\,\mathrm{kpc}<\lambda<20\,\mathrm{kpc}$. 

The simulations adopt a cubic box of size $2^{28}\,r_\mathrm{g}\approx2.7\times10^8\,r_\mathrm{g}$ ($\approx80\,\mathrm{kpc}$ for M87) in each direction and cover a radial domain of $[1,2^{27}(1.3\times10^8)]\,r_\mathrm{g}$ ($[0.3\,\mathrm{mpc},40\,\mathrm{kpc}]$). The mesh is similar to \citet{Guo2023ApJ...946...26G} but the root grid is a cube of $256^3$ cells, double the resolution of previous hydrodynamic runs. To alleviate the excessively severe time-step restriction on small scales, we first evolve the simulations with fewer levels of mesh refinement and a larger inner radius $r_\mathrm{in,\,old}$. Then we restart the simulation, add 4 more levels of mesh refinement, and set a smaller inner radius $r_\mathrm{in,\,new}=\frac{1}{16}r_\mathrm{in,\,old}$. The new cells in the simulation domain are initialized with the values used for the boundary conditions. The duration of each run is at least several $10^4$ dynamical times at the corresponding inner boundary. In this way, we successively zoom in to the smallest scale. In the fiducial suite of runs, we use 7, 11, 15, 19, and 23 levels of mesh refinement with the highest resolution $\Delta x=1/8\,r_\mathrm{g}$ and set $r_\text{in}=32768\,(2^{15}),2048\,(2^{11}),128,8, \text{and}\, 1\,r_\text{g}$, respectively. Each level is resolved with $256^3$ cells so the whole domain is resolved by $\sim 4\times10^8$ cells. We also convert the final MHD run to a GRMHD run with a black hole spin $a=0.9375$. The results of this GRMHD run will be presented in the future. For data analysis, we focus on scales $r\geq8\,r_\text{g}$ of the MHD runs.

We fix the outer boundary (the cells with $r>2^{27}\, r_\mathrm{g}$) to be the initial conditions. The fixed outer boundary conditions have negligible effects on the accretion flow because of the large separation of spatial and time scales. For the inner boundaries ($r<r_\mathrm{in}$), we adopt a vacuum sink by resetting a fixed density $n_\mathrm{sink}=5\times10^{-3}\,\mathrm{cm^{-3}}$, temperature $T_\mathrm{sink}=6\times10^4\,\mathrm{K}$, and zero velocity every time step. Then to avoid a large Alfv\'en speed $v_{A}\equiv B^2/\rho$ within the sink, we apply a ceiling $v_{A,\mathrm{ceil}}\approx4v_\mathrm{K}(r_\mathrm{in})$ by increasing density without changing the magnetic field, where $v_\mathrm{K}(r)\equiv\sqrt{GM(<r)/r}$ is the local Keplerian velocity.

For safety, we typically use a Courant number of $0.2$ and a radius-dependent density floor of $10^{-3} n_\mathrm{init}(r=r_\mathrm{in}) (r/r_\mathrm{in})^{-3/2}$. We apply a temperature floor of $T_\mathrm{floor}=2\times10^{4}\,\mathrm{K}$ in the fiducial set of runs, lower than that in \citet{Guo2023ApJ...946...26G}, not only due to the double resolution but also because the magnetic field can increase the scale height of the cold disk, thus alleviating the requirement of higher temperature floors.

Each run costs about 2000 GPU hours and the fiducial set of runs costs 10,000 GPU hours in total.

\begin{figure*}[ht]
    \centering
    \vspace{-1mm}
    \includegraphics[width=0.92\linewidth]{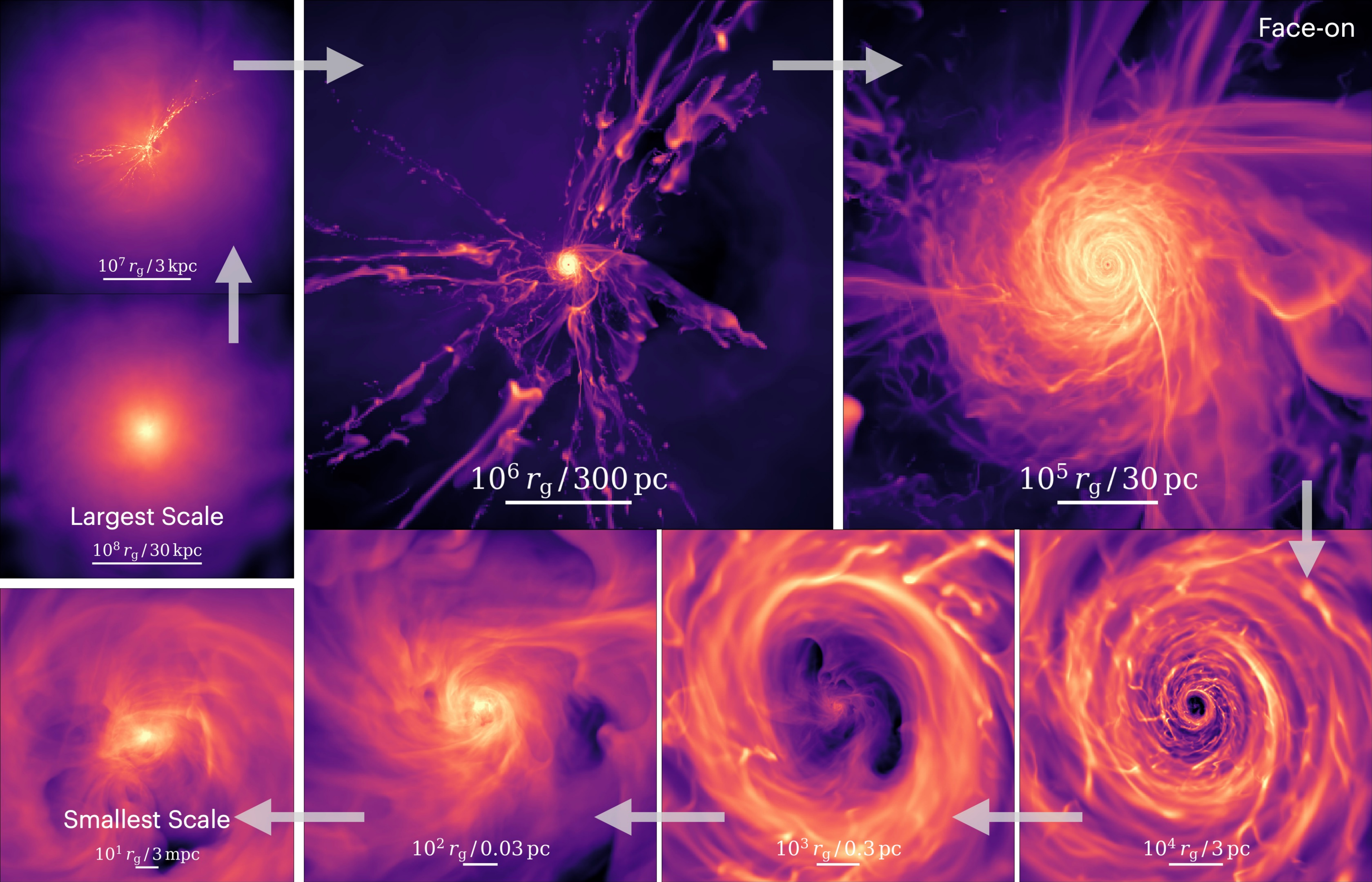}\\
    \vspace{1.5mm}
    \includegraphics[width=0.92\linewidth]{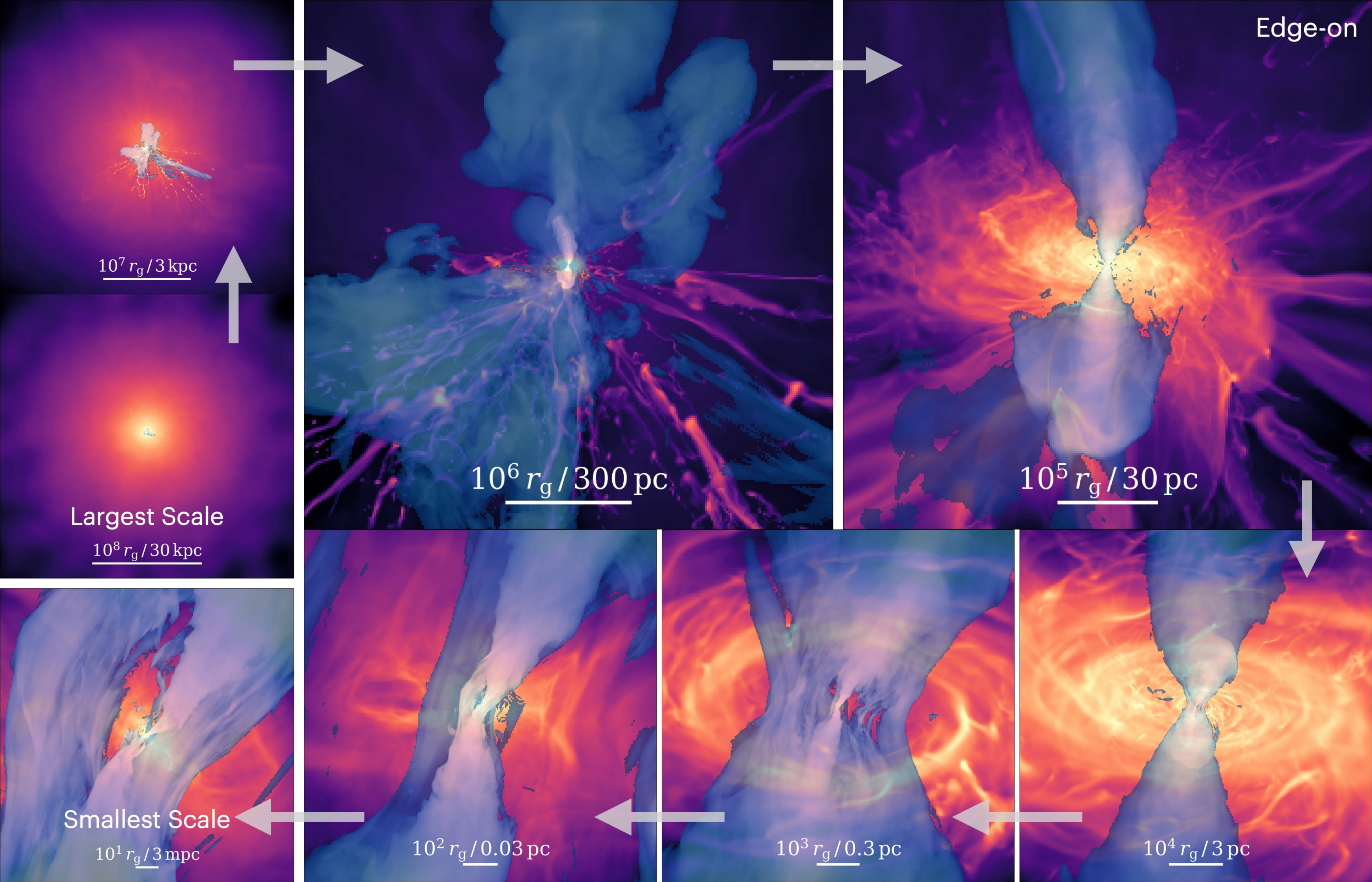}\\
    \vspace{-1mm}
    \caption{Series of images of the projected gas density on a logarithmic scale from black to white in our fiducial suite of simulations; we show both face-on (top) and (slightly tilted) edge-on (bottom) views on different radial scales. The viewing orientation is the same in every panel. Each panel rescales to its dynamic range, from $N\sim 10^{21}\,\mathrm{cm^{-2}}$ ($n\sim 10^{-2}\,\mathrm{cm^{-3}}$) on the largest scale to $N\sim 10^{23}\,\mathrm{cm^{-2}}$ ($n\sim 10^{6}\,\mathrm{cm^{-3}}$) on the smallest scale. Projected Bernoulli parameter density $\rho\mathcal{B}$ (see \eq\ref{eq:Be}) for gas with $\mathcal{B}>4T_\mathrm{init}$ is overplotted on a logarithmic scale from dark blue to light blue in the edge-on view to highlight the outflow. Structures include hot virialized gas, multi-phase chaotic inflow, ordered turbulent cold disk and energetic hot outflow, and MAD-like hot accretion flow from large to small scales.}
    \label{fig:zoom}
\end{figure*}

\begin{figure*}[ht]
    \centering
    \includegraphics[width=\linewidth]{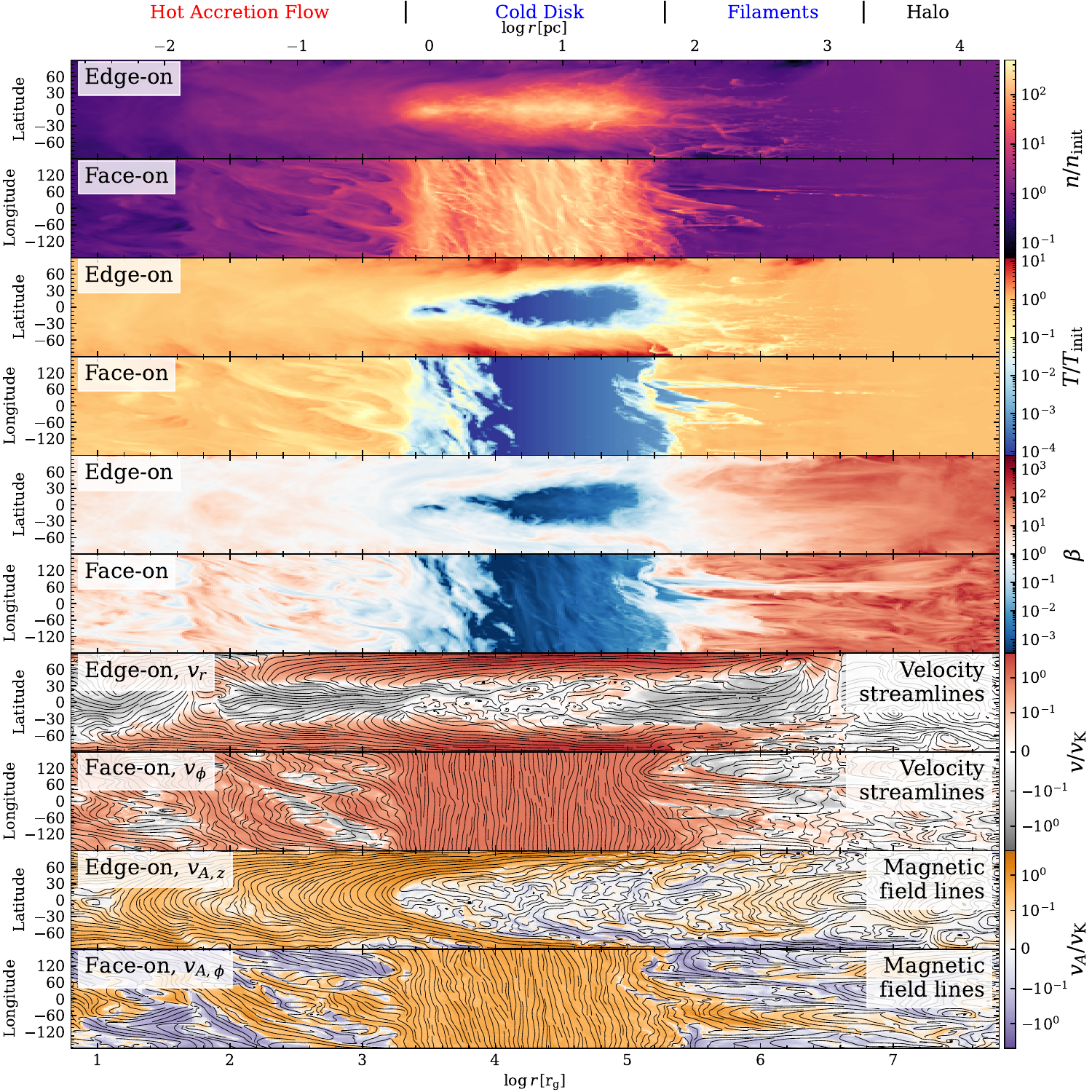}
    \caption{Images of (from top to bottom) gas density, temperature, plasma $\beta$, radial and rotational velocity, and radial and rotational Alfv\'en velocity on a logarithmic scale. Each panel shows the azimuthally averaged radius-latitude images (face-on) in the upper subpanel and vertically (for $|\cos\theta|<0.2$) averaged radius-longitude images (edge-on) in the lower subpanel. Density and temperature are normalized by the hydrostatic initial profile $n_\mathrm{init}(r)$ and $T_\mathrm{init}(r)$. Velocities are normalized by the local Keplerian velocity $v_\mathrm{K}(r)$. 
    The gas on scales $\gtrsim 3\,\mathrm{kpc}$ ($10^7\,r_\mathrm{g}$) is virialized ($T\sim T_\mathrm{init}$), diffuse, nearly spherically symmetric, and weakly magnetized with $\beta\sim 100$.
    The cold filaments on scales $\sim 0.03-3\,\mathrm{kpc}$ ($10^5-10^7\,r_\mathrm{g}$) are chaotic with density $\sim$ 100 times of the hot gas, magnetized with $\beta\sim 1$, and velocities of nearly free-fall speed.
    On scales $\sim 0.3-30$ pc ($10^3-10^5\,r_\mathrm{g}$), the disk is cold ($T\sim 10^{-4} T_\mathrm{init}$), Keplerian ($v_\phi\approx v_\mathrm{K}$), geometrically thick ($H/r\sim0.5$), and highly magnetized ($\beta\sim10^{-3}$) with primarily toroidal magnetic flux. The outflow is hot ($T\gtrsim T_\mathrm{init}$), magnetized ($\beta\sim1$) with primarily poloidal flux, super-Keplerian, and concentrated towards the poles with a half-opening angle $\lesssim 30^\circ$. 
    The cold disk is truncated around $\lesssim 0.3\,\mathrm{pc}$ ($10^3\,r_\mathrm{g}$) and transitions to a hot ($T\sim T_\mathrm{init}$) turbulent disk-like accretion flow. The accretion is essentially in a MAD state with a turbulent magnetic field of $\beta\sim1$ and a considerable coherent vertical magnetic flux. The mean flow pattern is inflow in the midplane and outflow in the polar region from the scale of filaments to the hot accretion flow.
    \label{fig:logr}}
\end{figure*}

\begin{figure*}[ht]
    \includegraphics[width=\linewidth]{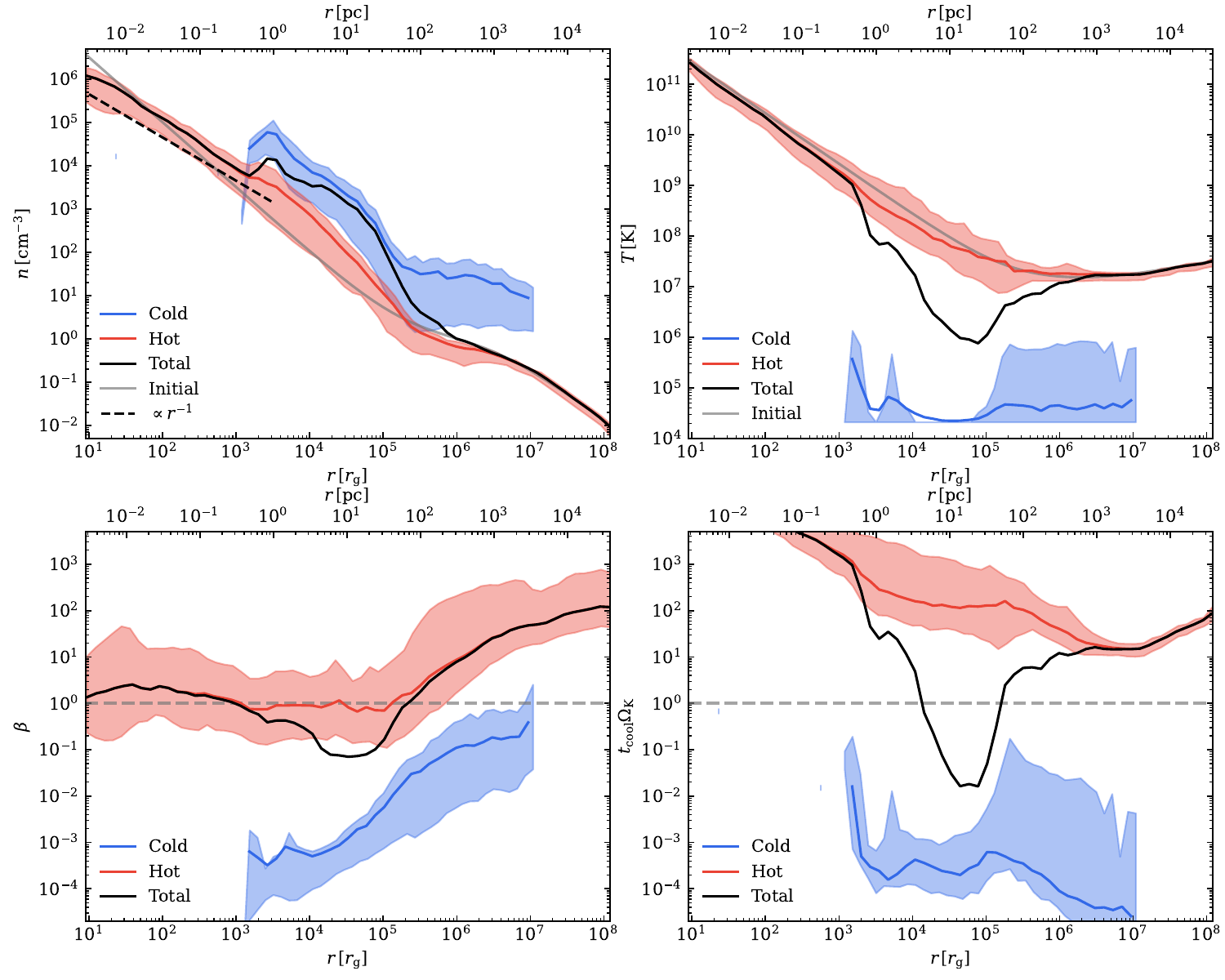}
    \caption{Radial profiles of several properties versus radius. Blue shows cold gas, red shows hot gas, and black shows total. Lines show the mean and the shaded ranges show 10\% to 90\% volume inclusion interval. Top left: volume-weighted number density. The partially transparent grey line shows the initial hydrostatic equilibrium profile. The cold gas is denser by a factor of $\sim 100$. Top right: mass-weighted temperature with the partially transparent grey line showing the initial profile. Bottom left: plasma $\beta$, the average value is defined by $\bar{\beta}=\langle{P}_\mathrm{gas}\rangle/\langle{P}_\mathrm{mag}\rangle$, i.e., the ratio of volume-weighted thermal pressure and magnetic pressure. The cold gas is increasingly magnetized at smaller radii, reaching $\beta\sim 10^{-4}$ at $r\sim 10^3\,r_\mathrm{g}$. The hot gas saturates to $\beta\sim1$ within $\sim30\,\mathrm{pc}$ ($10^5\,r_\mathrm{g}$). Bottom right: cooling time scales $t_\mathrm{cool}$ normalized by $\Omega_\mathrm{K}\equiv\sqrt{GM(<r)/r^3}$. The mean value is defined by the total internal energy divided by the total cooling rate in the radial shell. The cooling time is very short for cold gas ($t_\mathrm{cool}\Omega_\mathrm{K}\lesssim 10^{-3}$) but long for hot gas ($t_\mathrm{cool}\Omega_\mathrm{K}\gtrsim 10^{2}$). The highly magnetized cold gas transitions to a hot flow (\figu\ref{fig:scalar}) in spite of its short cooling time, likely due to strong magnetic heating via mechanisms like reconnection in the $\beta \ll 1$ gas.
    \label{fig:rad}}
\end{figure*}

\begin{figure*}[ht]
    \includegraphics[width=\linewidth]{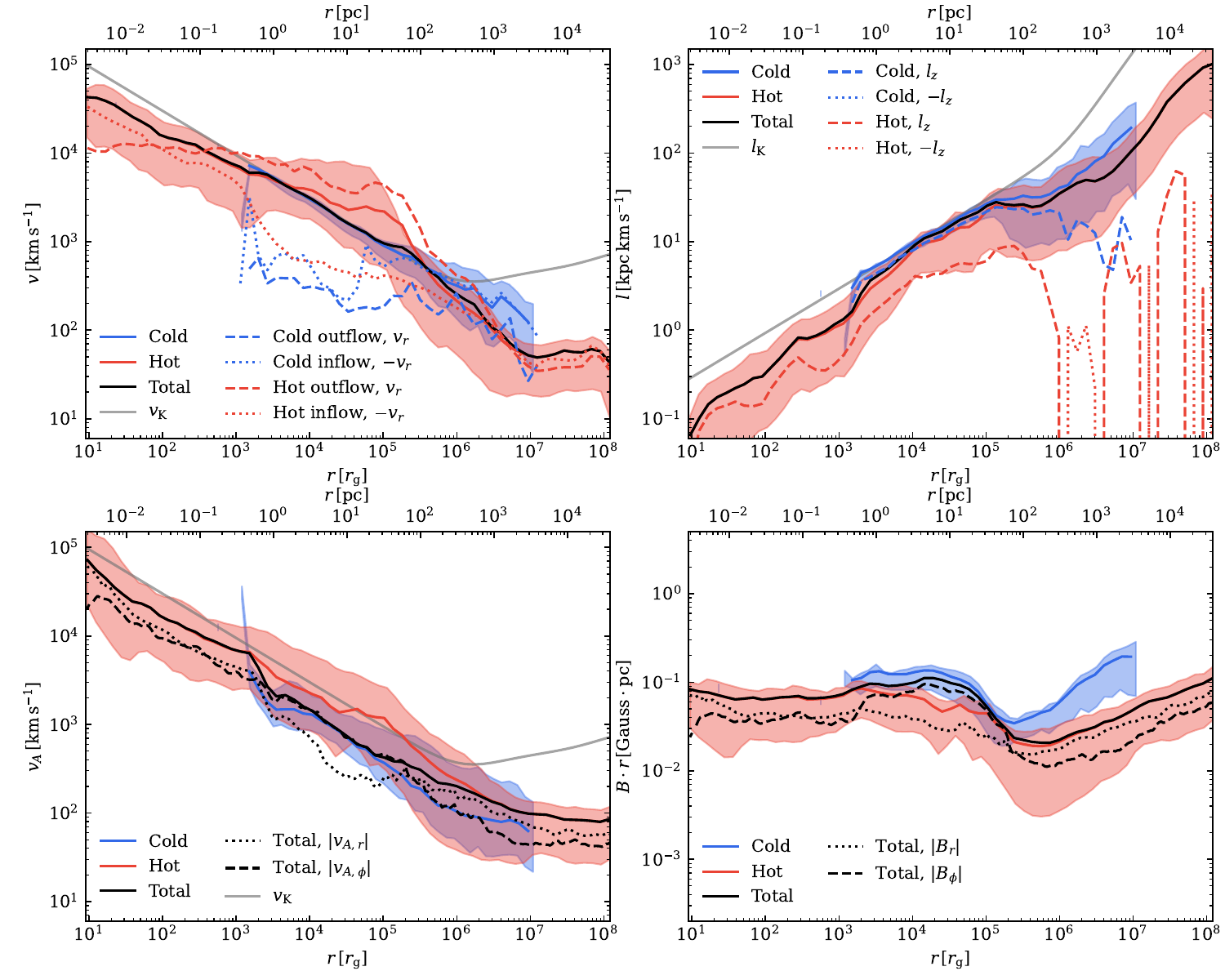}
    \caption{Similar to \figu\ref{fig:rad}, but for other properties. Top left: mass-weighted speed of cold, hot, and total gas and its radial component, $v_r$, for cold/hot outflow/inflow. The partially transparent grey line marks the Keplerian speed. Top right: mass-weighted specific angular momentum $l=r\sqrt{v^2-v_r^2}$ for cold, hot, and total gas and its $z$ component $l_z=(xv_y-yv_x)$ for cold and hot gas. The partially transparent grey line shows the Keplerian angular momentum $l_\mathrm{K}\equiv rv_\mathrm{K}$. Bottom left: mass-weighted Alfv\'en speed and its components $|v_{A,r}|$ and $|v_{A,\phi}|$ with the partially transparent grey line marking the Keplerian speed. Bottom right: volume-weighted magnetic field strength times radius (to reduce the dynamic range) and its components $|B_r|$ and $|B_\phi|$. The cold filaments are nearly free-fall, the cold disk is rotation-dominated, and the hot gas has an outflow speed of $\sim 10^4\,\mathrm{km\,s^{-1}}\gtrsim 5v_\mathrm{K}$ on scales $\sim 10\,\mathrm{pc}$, corresponding to the strong outflow. Most gas has a high Alfv\'en speed $v_{A}$ close to $v_\mathrm{K}$ inside $\sim100\,\mathrm{pc}$. The magnetic field roughly follows $B\sim r^{-1}$. The magnetic field is primarily toroidal on the cold disk scale and more poloidal on the hot flow scale.
    \label{fig:radv}}
\end{figure*}

\begin{figure*}[ht]
    \includegraphics[width=\linewidth]{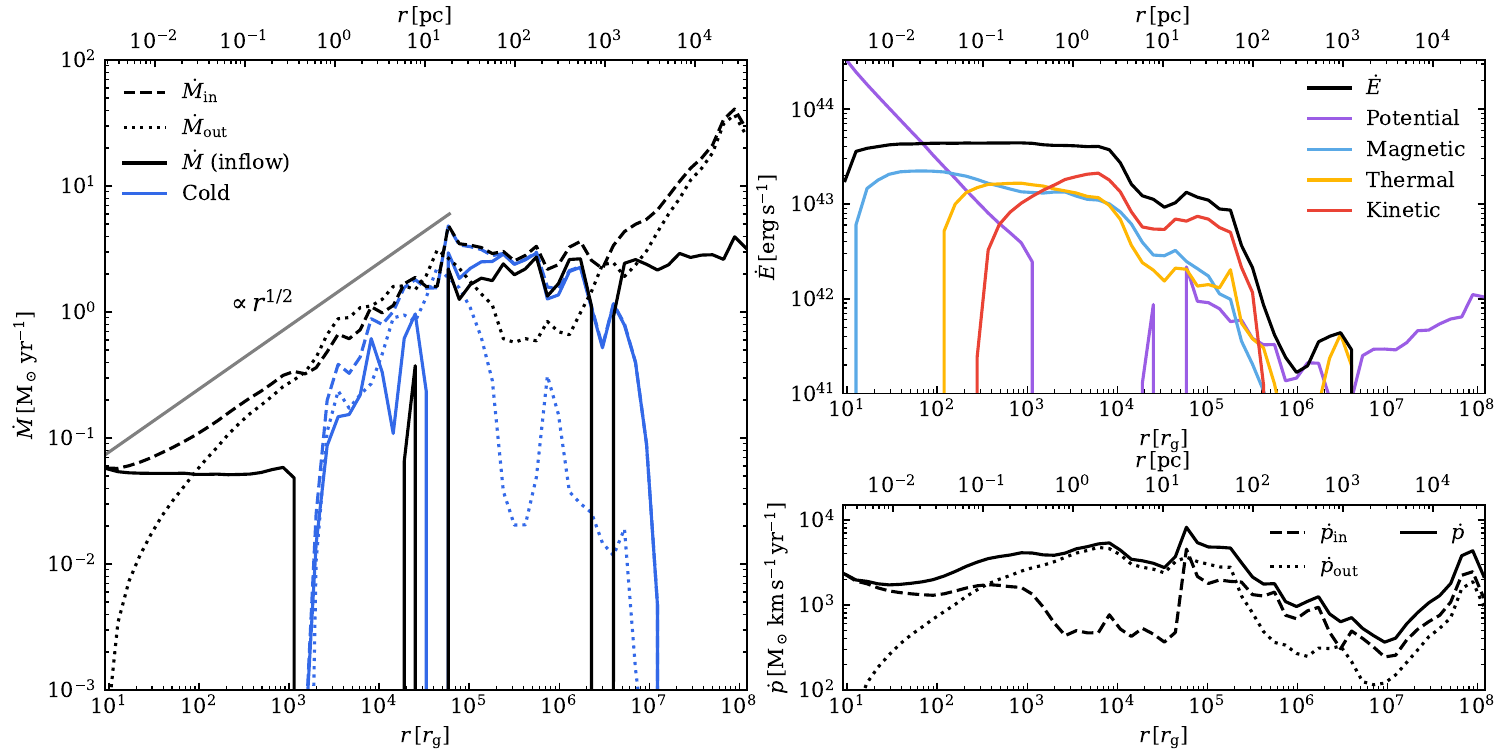}
    \caption{Left: mass inflow, outflow, and net flow $\dot{M}=\dot{M}_\mathrm{in}-\dot{M}_\mathrm{out}$ (see \eq\ref{eq:mdot}). There are strong inflow and outflow but only a small net flow toward the sink region. The accretion is dominated by cold gas for $r>10^3\,r_\mathrm{g}$ and hot gas for $r<10^3\,r_\mathrm{g}$. Top right: energy flow including the potential, magnetic, thermal, and kinetic components (see \eq\ref{eq:edot}). There is a strong hot outflow in the inner 100 pc. Magnetic, thermal, and kinetic energy successively dominate the energy outflow from smaller to larger scales. Bottom right: radial momentum flow. It is $\sim 3\times10^3\, M_\odot\,\mathrm{km\,s^{-1}\,yr^{-1}}$, essentially independent of radius. At large radii, the net flow is not independent of radii because we cannot run long enough and the solutions here are \textit{not} statistical steady state. They may be at some intermediate radii but even that is not totally clear.
    \label{fig:rad_flow}}
\end{figure*}

\begin{figure*}[ht]
    \centering
    \includegraphics[width=\linewidth]{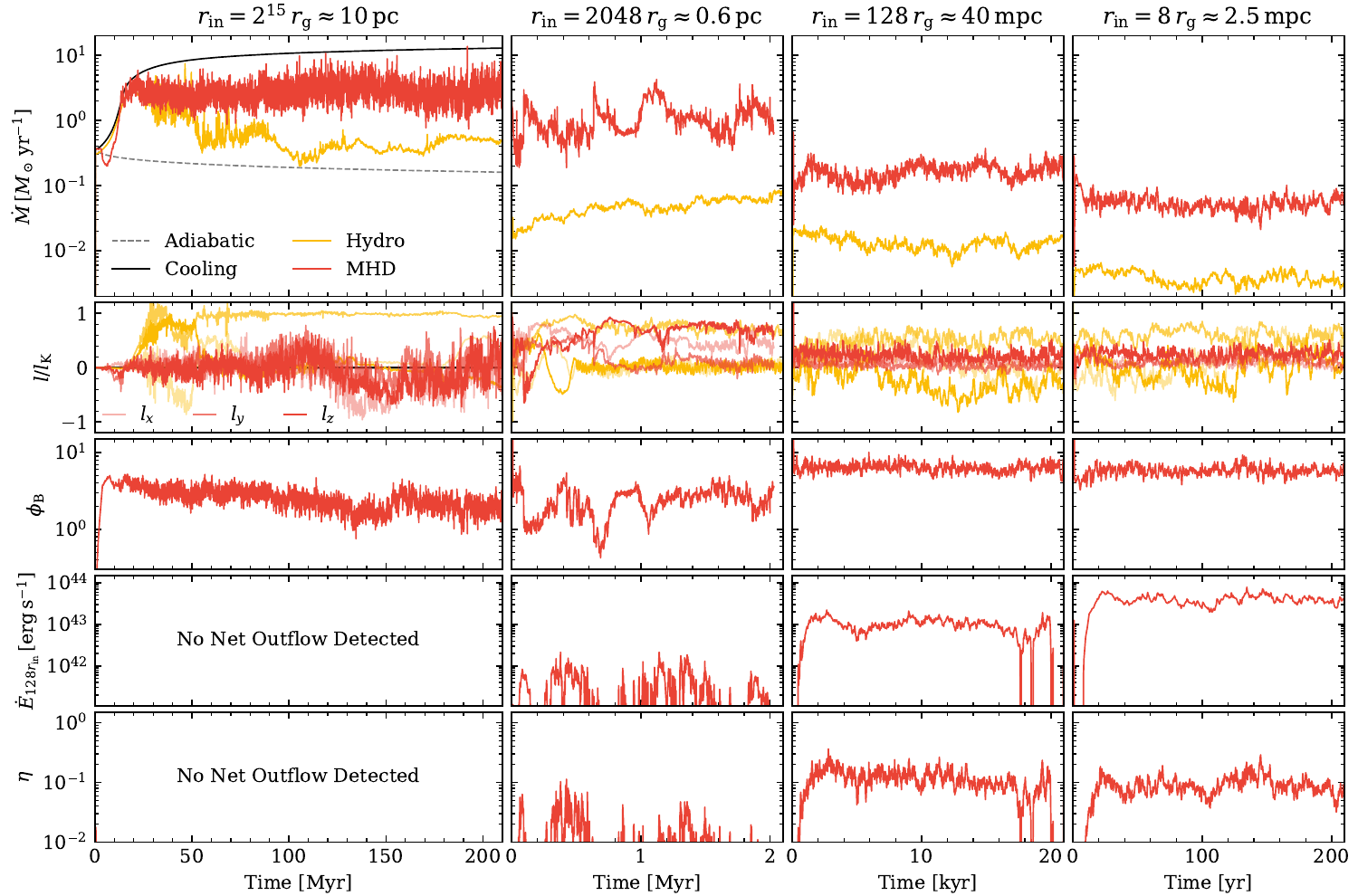}
    \caption{Time evolution of (from top to bottom) mass-accretion rate, three components of mass-weighted angle-averaged specific gas angular momentum through $r_\mathrm{in}$ normalized by local specific Keplerian angular momentum, normalized magnetic flux (see \eq\ref{eq:phi}), net energy flux measured at $128\,r_\mathrm{in}$ (\eq\ref{eq:edot} but without potential energy), and feedback efficiency $\eta$ (see \eq\ref{eq:eta}) for different inner radii $r_\mathrm{in}$. The mass accretion rate and angular momentum for a hydrodynamic run are plotted in blue for reference. The energy flux for the hydrodynamic run is negative so not shown here. During most time and scales of the simulations, the accretion flow has a higher mass accretion rate and lower angular momentum than the hydrodynamic run. The magnetic flux saturates to a MAD state ($\phi_\mathrm{B}\sim 6-10$) on small scales. With smaller $r_\mathrm{in}$, the mass accretion rate decreases but energy output increases, with an approximately constant feedback efficiency $\eta$.
    \label{fig:evo}}
\end{figure*}

\begin{figure*}[ht]
    \includegraphics[width=\linewidth]{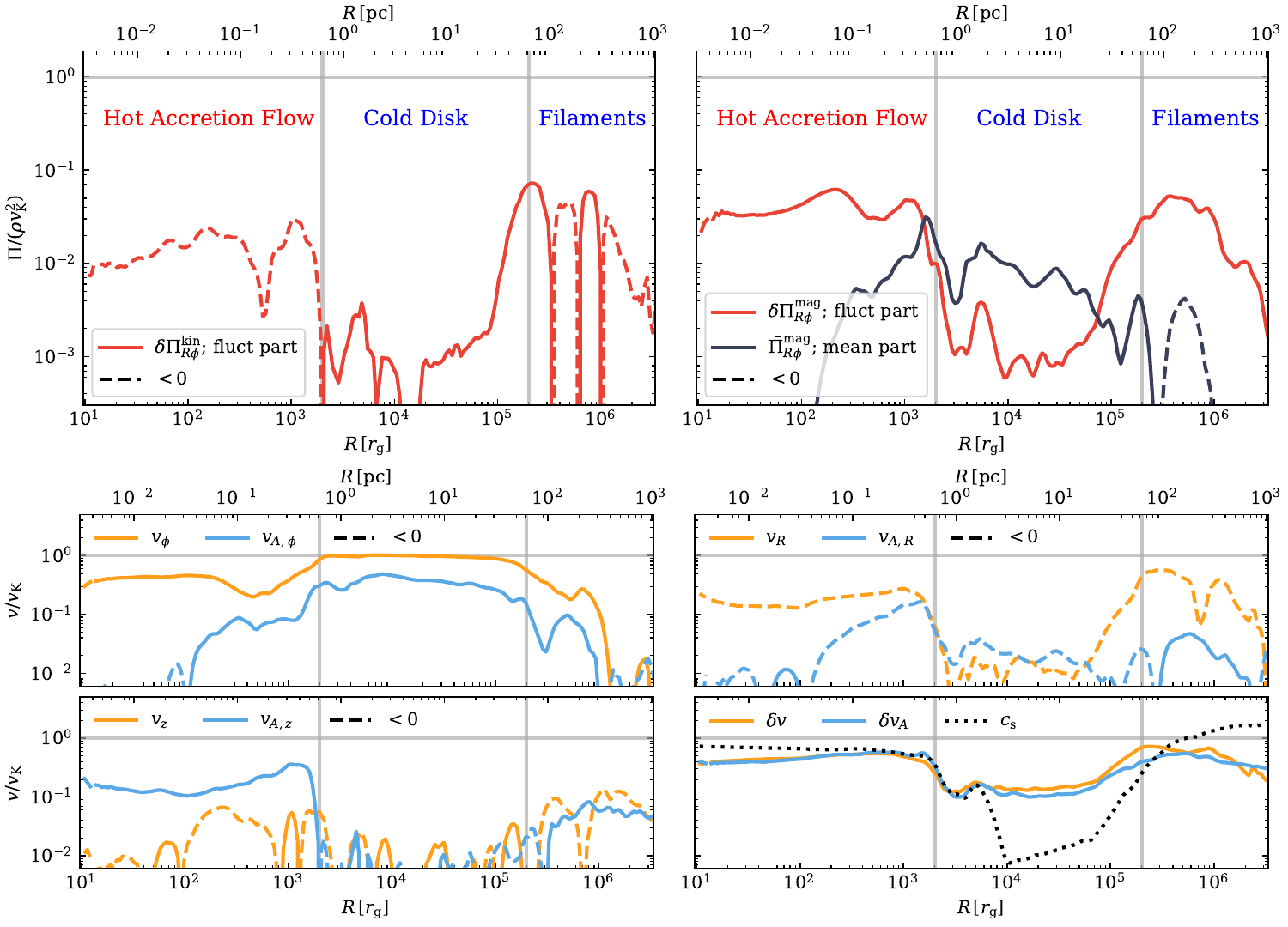}
    \caption{Time and angle-averaged profiles of mean and fluctuating stresses, velocities, and Alfv\'en velocities in the midplane (averaged over $|z/R|<0.2$) versus cylindrical radius. The stresses are normalized by $\rho v_\mathrm{K}^2$ while the speeds are normalized by $v_\mathrm{K}$. Solid (dashed) lines correspond to positive (negative) values. Though not shown, the three components of the turbulent velocity and turbulent Alfv\'en velocities are similar to order of magnitude. Strong magnetic and kinetic stresses drive angular momentum transfer on all scales shown here. The accretion flow is highly turbulent with $\delta v\sim \delta v_{A}\sim v_\mathrm{K}$. The cold disk is Keplerian with mean inflow velocity $v_R\sim -0.01 v_\mathrm{K}$. The hot accretion flow is thermally supported with large mean inflow $v_\mathrm{R}\sim -0.1 v_\mathrm{K}$ and a coherent vertical magnetic flux crossing the midplane.
    \label{fig:disk_radial}}
\end{figure*}

\begin{figure*}[ht]
    \includegraphics[width=\linewidth]{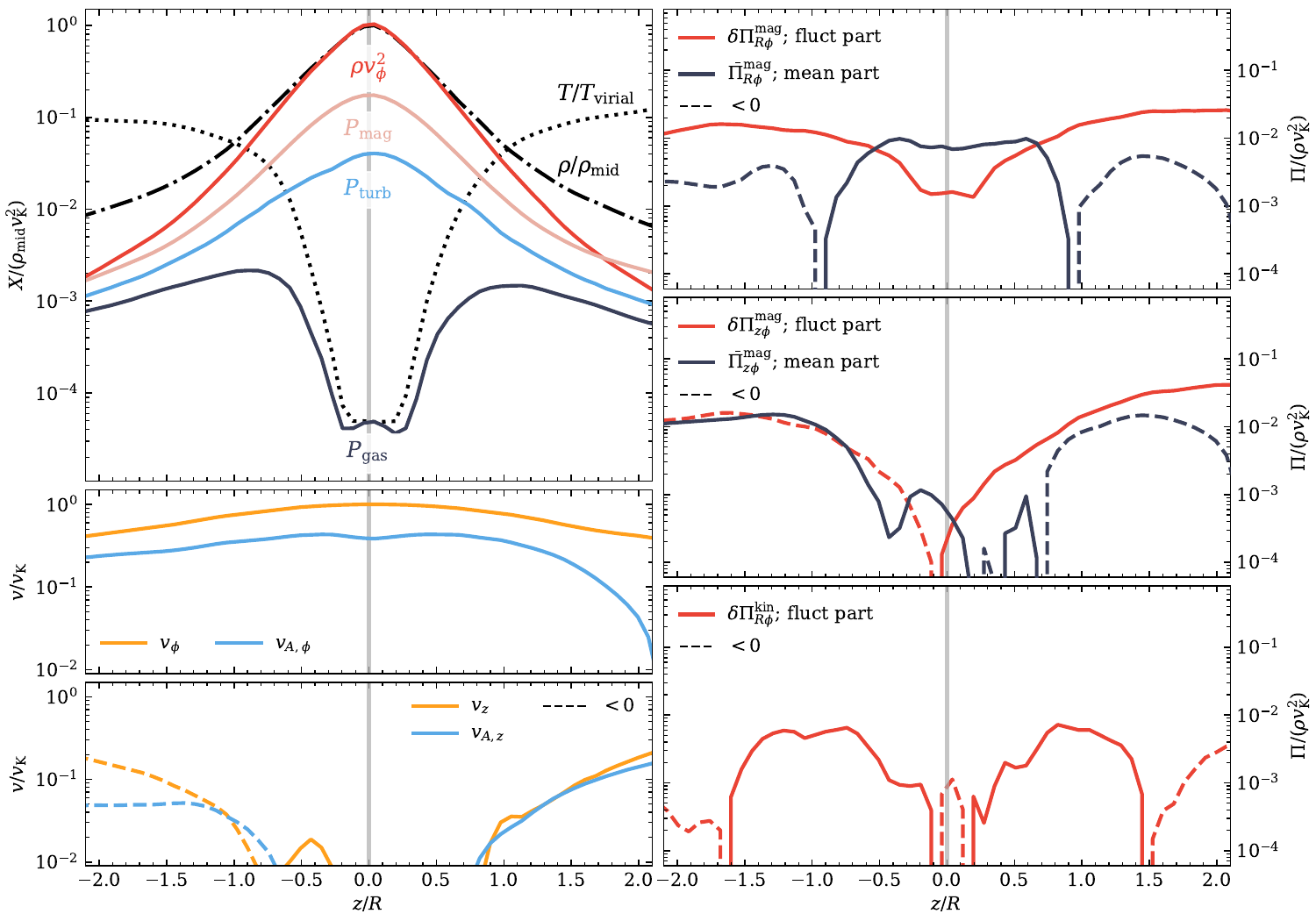}
    \caption{Vertical profiles of various time and azimuthally averaged properties in the cold magnetized disk (see \figu\ref{fig:logr}), corresponding to a radial shell with $10^4<R/r_\mathrm{g}<3\times10^4$, i.e., $3-10\,\mathrm{pc}$. Top left: rotational velocity, magnetic pressure, turbulent pressure, and thermal pressure with density and temperature as reference. Middle left: angular velocity and toroidal Alfv\'en velocity. Bottom left: vertical velocity and Alfv\'en velocity. The velocities are normalized by $v_\mathrm{K}$ while the stresses are normalized by $\rho v_\mathrm{K}^2$. Top right: $R\phi$-component of Maxwell stress. Middle right: $z\phi$-component of Maxwell stress (the wind stress). Bottom right: $R\phi$-component of Reynolds stresses. Solid (dashed) lines correspond to positive (negative) values. The cold disk is thick, dominated by toroidal magnetic fields, and highly turbulent with strong magnetic and kinetic stresses driving angular momentum transfer and outflow.
    \label{fig:disk_vertical}}
\end{figure*}

\begin{figure*}[ht]
    \includegraphics[width=\linewidth]{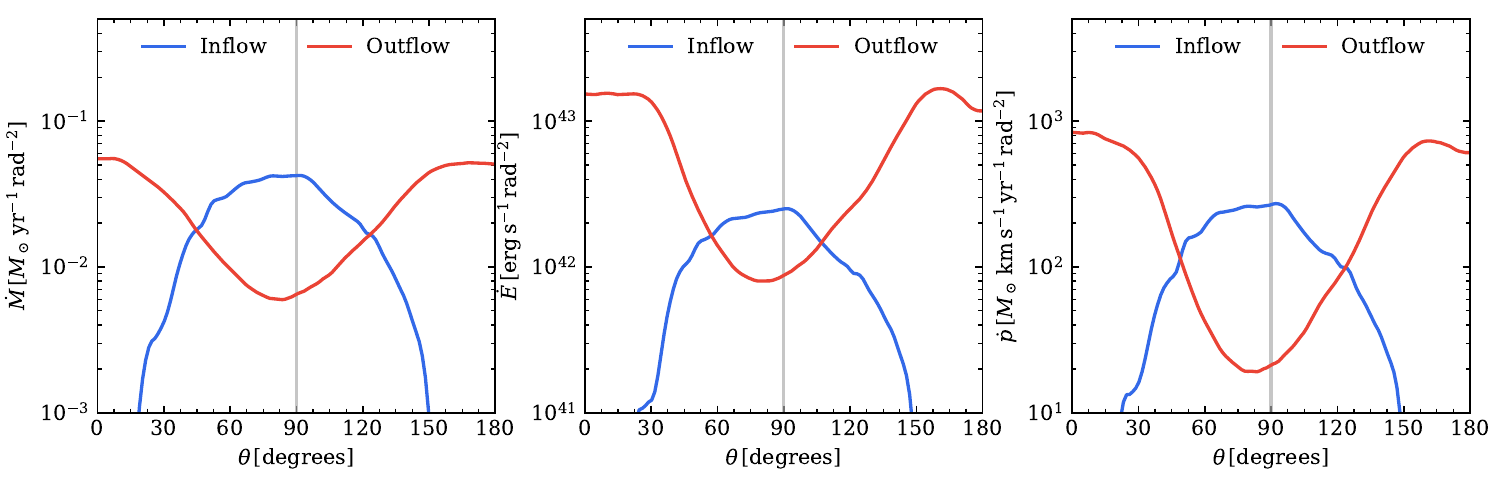}
    \caption{Angular profile of mass flux (left), energy flux (without potential energy) (middle), and momentum flux (right) averaged in the radial shell of $5\times10^2-2\times10^3\,r_\mathrm{g}$ ($\approx 0.15-0.6\,\mathrm{pc}$), where the flow is primarily hot (\figu\ref{fig:logr}). Mass and energy inflows are in the midplane ($\sim 60^\circ-120^\circ$) while outflows are in the polar region ($\lesssim 30^\circ$ and $\gtrsim 150^\circ$) with mixing in between.
    \label{fig:theta_flux}}
\end{figure*}

\begin{figure*}[ht]
    \includegraphics[width=\linewidth]{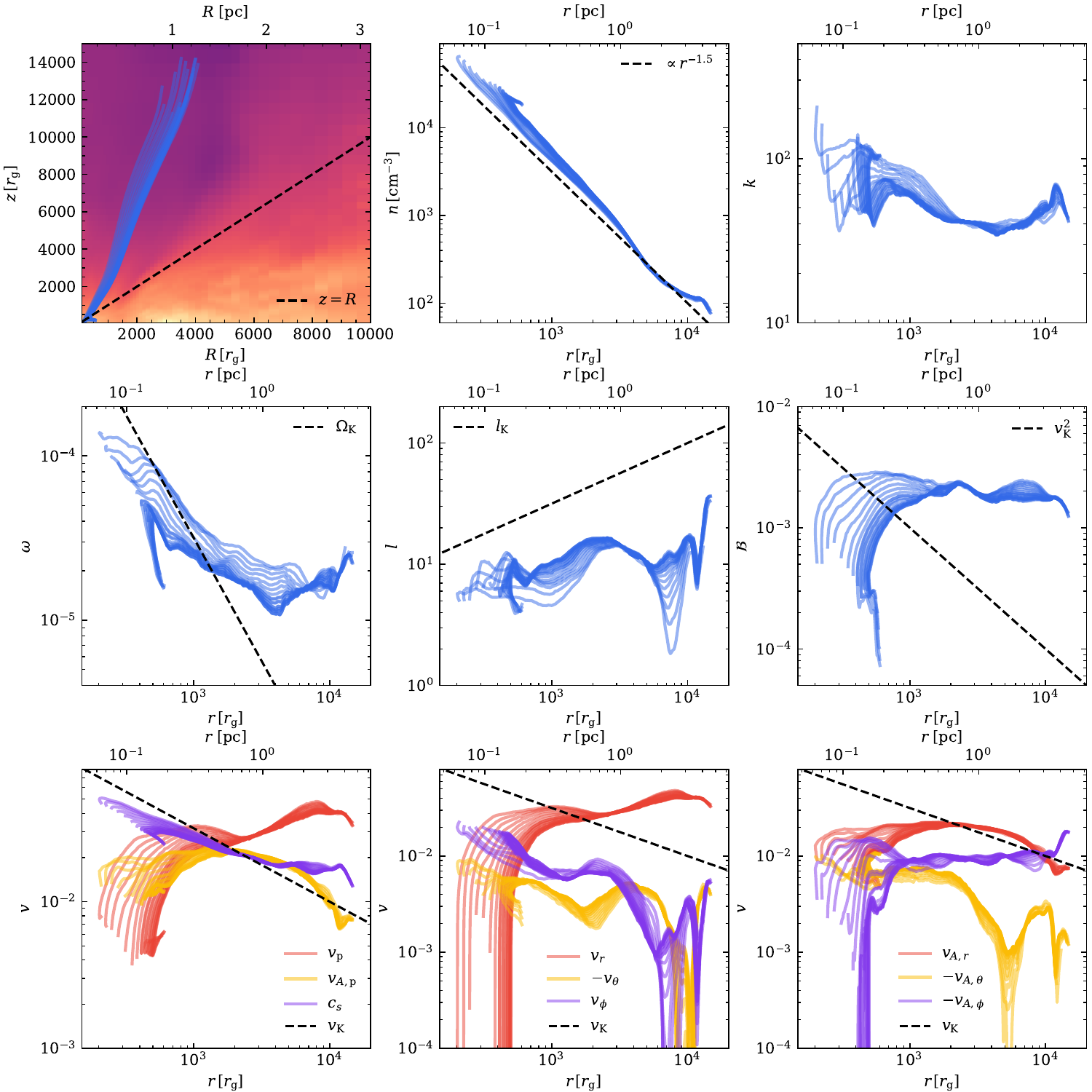}
    \caption{Variables along a series of streamlines of the outflow/wind. The units are in code units when not specified. Top left: the streamlines in the $R-z$ plane overplotted on the image of azimuthally averaged density. Top center: the number density versus spherical radius $r$. Top right: mass loading parameter $k$ (\eq\ref{eq:k}). Middle left: angular velocity $\omega$ (\eq\ref{eq:omega}). Middle center: angular momentum $l$ (\eq\ref{eq:l}). Middle right: the Bernoulli parameter $\mathcal{B}$ (\eq\ref{eq:Be}). Bottom left: poloidal speed, poloidal Alfv\'en speed, and sound speed. Bottom center: three components of the velocity. Bottom right: three components of Alfv\'en velocity. 
    The outflow/wind is super-Keplerian, supersonic, and super-Alfv\'enic. The conserved quantities ($k,\omega,l,\mathcal{B}$) are roughly constant along streamlines for $r\gtrsim 10^3\,r_\mathrm{g}$.
    \label{fig:disk_wind}}
\end{figure*}

\begin{figure*}[ht]
    \includegraphics[width=\linewidth]{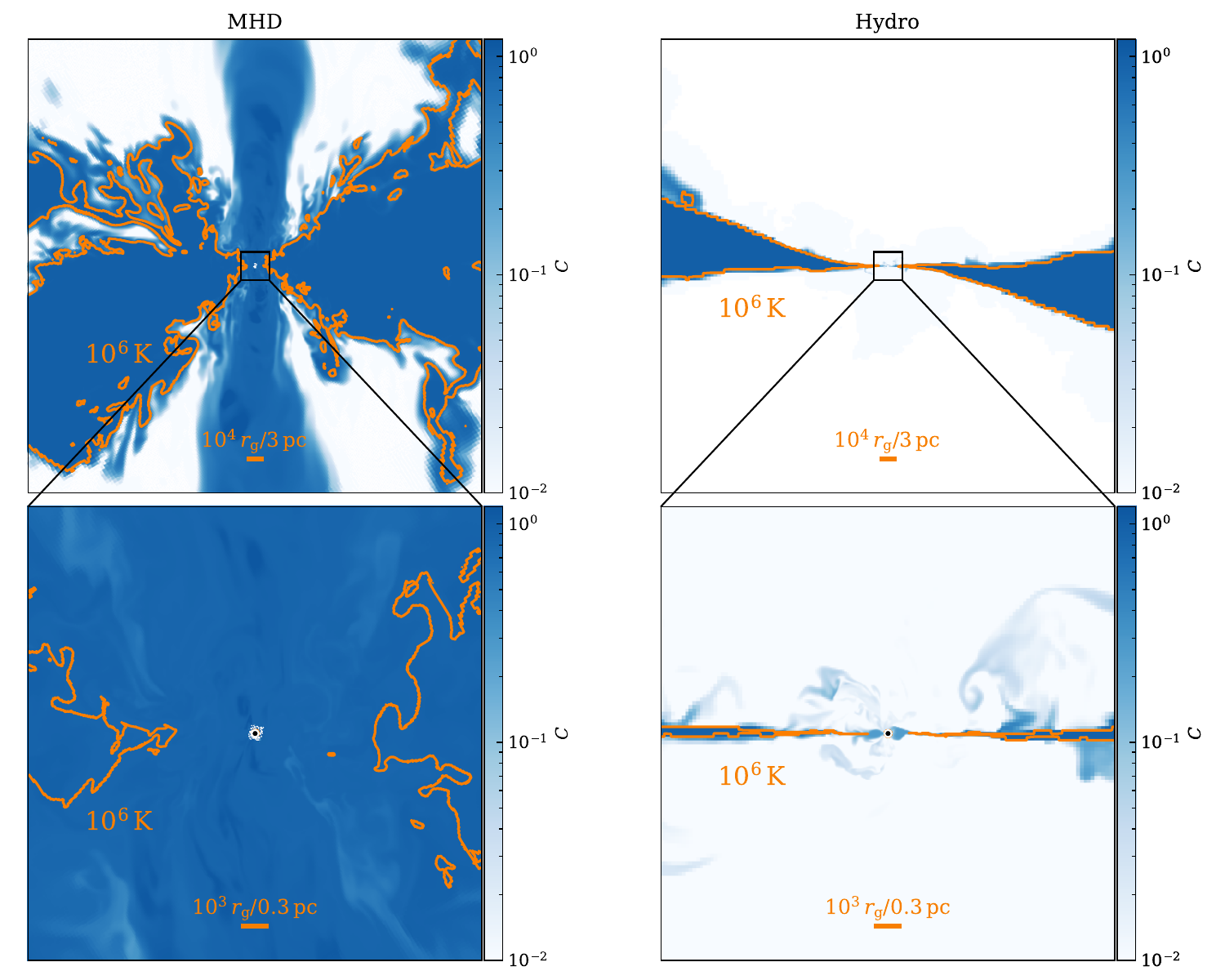}
    \caption{Slice of passive scalar fraction (\eq\ref{eq:scalar}) tracing the evolution of the cold gas for MHD run (left) and hydrodynamic run (right) for reference. The contours mark the temperature boundary between the cold disk and the hot accretion flow. A significant fraction ($\gtrsim 80\%$) of the hot accretion flow is from cold gas, as indicated by the large cold gas passive scalar fraction in the lower left panel. As a comparison, there is negligible mixing in the hydrodynamic case ($\lesssim 1\%$).
    \label{fig:scalar}}
\end{figure*}

\begin{figure*}[ht]
    \includegraphics[width=\linewidth]{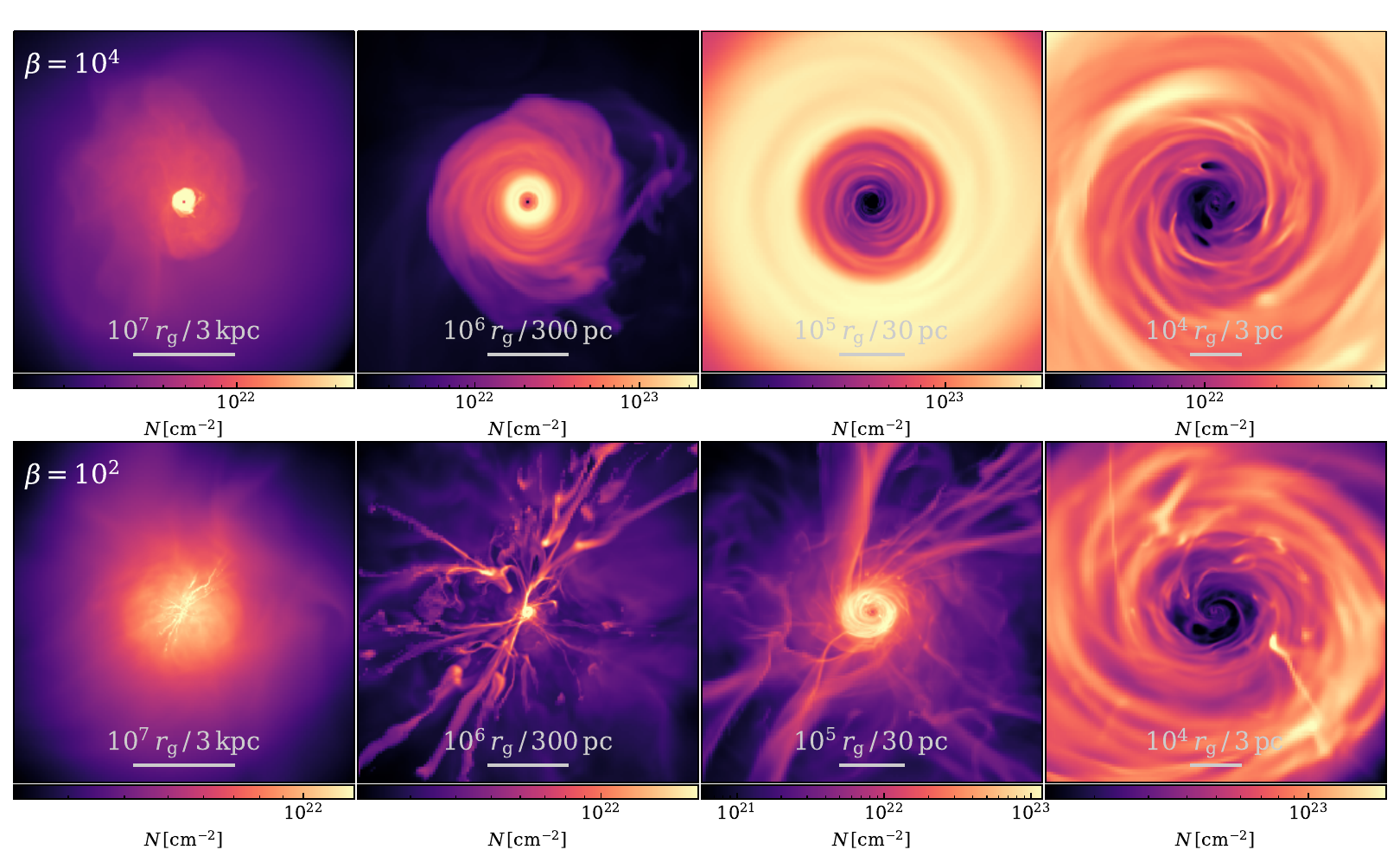}
    \caption{Similar to \figu\ref{fig:zoom} but for face-on views of projected density of two cases with half resolution of the fiducial run, temperature floor of $T_\mathrm{floor}=2\times10^5\,\mathrm{K}$, and different initial large-scale $\beta=10^4$ (top) and $\beta=10^2$ (bottom) on scales from $\sim 10\,\mathrm{kpc}$ (left) to $\sim 1\,\mathrm{pc}$ (right). When $\beta=10^4$, the accretion flow structure is similar to hydrodynamic cases; it has fewer filaments and forms a larger disk than the case with $\beta=10^2$. Note that the range of projected density varies from panel to panel.
    \label{fig:proj_beta}}
\end{figure*}

\begin{figure*}[ht]
    \includegraphics[width=\linewidth]{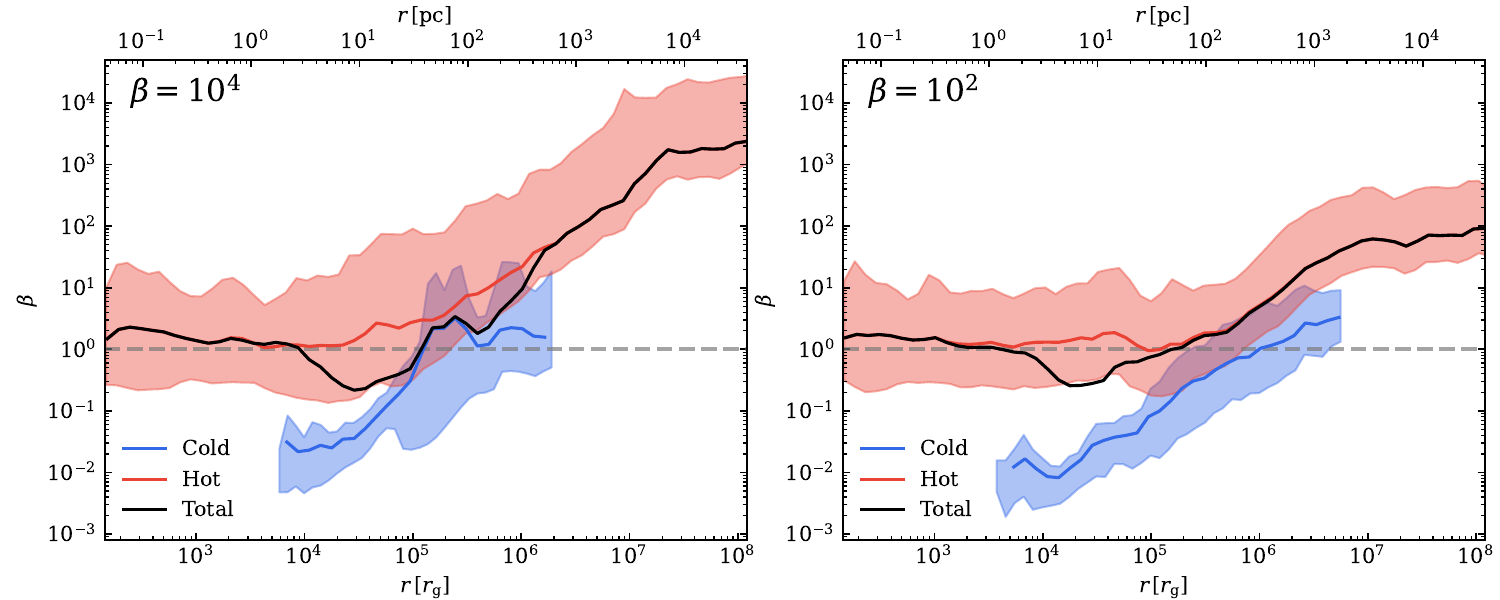}
    \caption{Similar to the bottom left panel in \figu\ref{fig:rad} showing radial profiles of plasma $\beta$, but for two cases shown in \figu\ref{fig:proj_beta} with initial large-scale $\beta=10^4$ (left) and $\beta=10^2$ (right). For $\beta=10^4$, the accretion flow is less magnetized than the run with $\beta=10^2$ on scales $\gtrsim 10^5\,r_\mathrm{g}$. In both cases, the magnetic field finally saturates within $\sim 10^4\,\mathrm{g}$. The cold gas is still highly magnetized but less than the fiducial case partially because the temperature floor here $T_\mathrm{floor}=2\times10^5\,\mathrm{K}$ is higher.
    \label{fig:rad_beta}}
\end{figure*}

\begin{figure}[ht]
    \includegraphics[width=\linewidth]{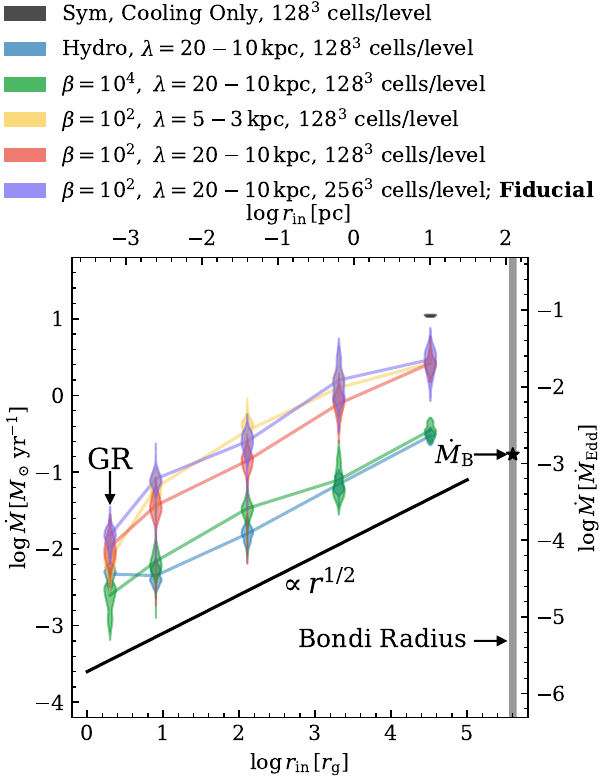}
    \caption{Relationship between mass accretion rate and inner radius for various runs. The normalization of the accretion increases with a stronger magnetic field (for initial $\beta\lesssim10^2$) but also depends on the exact stage of the accretion (ordered or chaotic). Overall, we find a universal scaling of $\dot{M}\propto r_\mathrm{in}^{1/2}$ over a large dynamic range from Bondi scale to horizon scale, spanning three to five orders of magnitude. In MHD, general relativistic effects of the spinning black hole lead to a somewhat lower accretion rate than would be predicted by extrapolating the Newtonian simulations from larger radii.
    \label{fig:scaling}}
\end{figure}

\section{Results} \label{sec:results}

\subsection{Basic Properties}

\figu\ref{fig:zoom} illustrates the gas morphology on a large dynamic range covering $\sim8$ orders of magnitude, from the circumgalactic medium on the galactic scale down to the accretion disk near the event horizon. As an alternative illustration, \figu\ref{fig:logr} plots the azimuthally-averaged and vertically-averaged images on a logarithmic scale using a new coordinate system $(r,\theta,\phi)$ defined by the orientation of the cold disk. Throughout the remainder of the paper, we define the gas with $T<T_\mathrm{cold}\equiv10\, T_\mathrm{floor}=2\times10^5\,\mathrm{K}$ as cold gas, gas with $T\ge T_\mathrm{hot}\equiv0.2\, T_\mathrm{init}(r)$ as hot gas, and gas in between ($T_\mathrm{cold}\le T< T_\mathrm{hot}$) as the intermediate gas. We note that, though we mainly discuss cold and hot gas, there is always a considerable fraction of intermediate gas in between. The basic properties from largest to smallest scales are as follows:
\begin{itemize}
    \item Diffuse Hot Halo: On scales $\gtrsim 3$ kpc ($10^7\,r_\mathrm{g}$), the gas is virialized ($T\sim 10^7\,\mathrm{K}$), diffuse ($n\sim 0.1\,\mathrm{cm^{-3}}$), nearly spherically symmetric, and weakly magnetized with $\beta\sim 100$ and $B\sim$ several $\mathrm{\mu G}$. The gas is in hydrostatic equilibrium with thermal pressure comparable to gravity and does not evolve significantly within the duration ($\lesssim 300\,\mathrm{Myr}$) of the simulations.
    \item Chaotic Cold Accretion along Magnetized Filaments: On scales $\sim 0.03-3$ kpc ($10^5-10^7\,r_\mathrm{g}$), the gas is multi-phase, composed of volume-filling hot gas, strong outflow from smaller scales, and chaotic cold ($T\lesssim 10^5\,\mathrm{K}$) inflow along filaments with densities $\sim$ 100 times of the hot gas, velocities of nearly free-fall speed, and magnetization of $\beta\sim1$ and $B\sim 100\,\mathrm{\mu G}$. The magnetic field within the cold gas is mostly aligned with the major axis of the filaments but sometimes bent significantly (\figu\ref{fig:logr}). The accretion flow is turbulent and magnetized, similar to the chaotic cold accretion~\citep{Gaspari2013MNRAS.432.3401G}.
    \item Cold Disk Accretion with Outflows: On scales $\sim 0.3-30$ pc ($10^3-10^5\,r_\mathrm{g}$), the cold gas is no longer chaotic but circularizes, forming a disk/torus with strong outflows. The disk is cold ($T\sim 10^4\,\mathrm{K}$), Keplerian ($v_\phi\approx v_\mathrm{K}$), geometrically thick ($H/R\sim0.5$), magnetically supported in the vertical direction $(v_{A}\sim 0.5v_\mathrm{K})$, highly magnetized ($\beta\sim10^{-3}$), and turbulent ($\delta v\sim0.3v_\mathrm{K}$) with strong spiral arms. The outflow is hot, magnetized, supersonic, super-Alfv\'enic, and super-Keplerian with $v_r\gtrsim v_A \sim c_\mathrm{s} \gtrsim v_\mathrm{K}$ and concentrated towards the poles with a half-opening angle $\lesssim 30^\circ$. The magnetic field is primarily toroidal in the disk midplane and poloidal in the polar region.
    \item Turbulent Hot Accretion Flow: On scales $\lesssim 0.3\,\mathrm{pc}$ ($10^3\,r_\mathrm{g}$), the cold disk is truncated and transitions to a hot ($T\sim T_\mathrm{init}$) turbulent disk-like accretion flow with $\delta v \sim v_\phi\sim 0.5 v_\mathrm{K}$. The accretion is essentially in a MAD state with a turbulent magnetic field of $\beta\sim1$ and a considerable coherent vertical flux.
\end{itemize}

The accretion flow is significantly different from the hydrodynamic case. First, the cold gas is more filamentary and chaotic on the scales of $\sim 300\,\mathrm{pc}$. The cold filaments lose their angular momentum more efficiently due to the magnetic tension and thus have a smaller circularization radius $r_\mathrm{circ}$, leading to a smaller disk. Second, the cold disk is puffier with scale height $H/R\sim v_{A}/v_\mathrm{K}\sim1$, instead of $H/R\sim c_\mathrm{s}/v_\mathrm{K}\ll1$ in hydrodynamic case. The details of disk structure and angular momentum transfer are analyzed in \sect\ref{subsec:disk}. Third, there is a super-Alfv\'enic and supersonic hot outflow driven by the magnetic field, different from the hydrodynamic case where the hot gas is either turbulent or a pure inflow. \sect\ref{subsec:outflow} analyzes the details of the outflow.

\figus\ref{fig:rad}, \ref{fig:radv}, and \ref{fig:rad_flow} more quantitatively examine the angle-averaged radial profiles of various quantities, including density, temperature, magnetic field, cooling time, velocities, mass flux, energy flux, and radial momentum flux for the multi-phase gas. The time and angle average of a variable $A$ is defined as
\begin{equation}
    \langle A \rangle \equiv \frac{\int_{t_0}^{t_1}\int_S A \dd\Omega \dd t}{(t_1-t_0)\int_S \dd\Omega} ,
\end{equation}
where $S$ is the area where we count the variable (the region of total, cold, and hot gas). The mass-weighted average is defined similarly by
\begin{equation}
    \langle A \rangle_m \equiv \frac{\langle \rho A \rangle}{\langle \rho \rangle}.
\end{equation}

The cold gas extends from $\sim 3\,\mathrm{kpc}$ ($10^7\,r_\mathrm{g}$) to $\sim 0.3\,\mathrm{pc}$ ($10^3\,r_\mathrm{g}$), with temperature $T\sim3\times10^4\mathrm{K}$ close to $T_\mathrm{floor}$, density higher than the background hot gas by a factor of $\sim 10-100$, and magnetic field stronger than that in the hot gas by a factor of a few to $\sim 10$. Due to the low temperature, the cold gas is highly and increasingly magnetized with decreasing radii, from $\beta\sim 1-0.1$ at $3\,\mathrm{kpc}$ to $\beta\sim 10^{-3}-10^{-4}$ at $0.3\,\mathrm{pc}$. The cooling time of the cold gas is much shorter than the local dynamical time with $t_\mathrm{cool}\Omega_\mathrm{K}\sim 10^{-2}-10^{-4}$, where $\Omega_\mathrm{K}\equiv v_\mathrm{K}/r=\sqrt{GM(<r)/r^3}$.

The hot gas temperature essentially remains similar to the initial conditions. Between $100\,\mathrm{pc}$ and $1\,\mathrm{pc}$, the density is higher than the initial profile with a steeper slope due to the mixing with the cold gas. Within $0.3\,\mathrm{pc} (10^3\,r_\mathrm{g})$, the accretion is again hot gas dominated so $\rho\sim r^{-1}$, similar to the turbulent hot gas accretion~\citep{Ressler2018MNRAS.478.3544R, Xu2019MNRAS.488.5162X, Ressler2020MNRAS.492.3272R, White2020ApJ...891...63W, Guo2020ApJ...901...39G, Guo2023ApJ...946...26G, Xu2023ApJ...954..180X}. The magnetic field is weak on large scales with $B\sim 1\,\mathrm{\mu G}$ and $\beta\sim100$, similar to the initial conditions. As radii decrease, the field strength gradually increases, reaching $\sim 10\,\mathrm{G}$ at $10\,r_\mathrm{g}$, with $\beta$ saturate to $\sim 1$ within $\sim 30\,\mathrm{pc}$. The hot accretion flow thus essentially enters the MAD state, similar to the fueling of Sagittarius A* \citep{Ressler2020MNRAS.492.3272R}.

Velocity, angular momentum, and magnetic field distribution of the multi-phase gas flow are shown in~\figu\ref{fig:radv}. The cold filaments on scales $\sim0.03-3\,\mathrm{kpc}$ are nearly free-fall, with radial velocity $v_r\sim v_\mathrm{K}$ dominating over other terms and radial Alfv\'en velocity $v_\mathrm{A,r}\sim 0.3v_\mathrm{K}$ larger than other components. In the disk region on scales $\sim0.3-30\mathrm{pc}$, the cold gas is instead rotation dominated with angular momentum $l\approx l_\mathrm{K}$. The thermal scale height of the disk $c_s/v_\mathrm{K}\sim 10^{-2}$ but the magnetic scale height $v_A/v_\mathrm{K}\sim 0.5$ with $v_{A,\phi}$ being the dominant term. The hot gas outflow is super-Keplerian, super-Alfv\'enic, and supersonic with velocity as large as $\sim 5\,v_\mathrm{K}$. The hot accretion flow within $\sim 10^{3}\,r_\mathrm{g}$ is again supported primarily by pressure instead of rotation with $c_\mathrm{s}\sim v_\mathrm{K}$ and $|v_r|\sim |v_\phi|\sim |v_{A,r}|\sim |v_{A,\phi}|\sim 0.3 v_\mathrm{K}$. The magnetic field follows $|B|\sim r^{-1}$ over a large dynamic range with $|B_\phi|$ dominating on the cold disk scale and $|B_r|\sim|B_\phi|$ on other scales.

\figu\ref{fig:rad_flow} shows the radial profiles of the mass flow, energy flow, and radial momentum flow. The mass flow rate defined by
\begin{equation}
\begin{aligned}
    \dot{M} \equiv & -\int\rho v_r r^2 \dd \Omega, \\
    = & \underbrace{\int_{v_r<0}\rho (-v_r) r^2 \dd \Omega}_{\dot{M}_\mathrm{in}}-\underbrace{\int_{v_r>0}\rho v_r r^2 \dd \Omega}_{\dot{M}_\mathrm{out}},\label{eq:mdot}
\end{aligned}
\end{equation}
includes both strong inflow ($\dot{M}_\mathrm{in}$) and outflow ($\dot{M}_\mathrm{out}$), leading to a small net mass inflow $\dot{M}=\dot{M}_\mathrm{in}-\dot{M}_\mathrm{out}$. The accretion is dominated by the cold gas for $r\gtrsim 10^3\,r_\mathrm{g}$ and by the hot gas after the transition ($r\lesssim 10^3\,r_\mathrm{g}$). Within $\sim 20\,\mathrm{pc}$, the accretion rate follows $\dot{M}_\mathrm{in}\propto r^{1/2}$. The energy flux is defined by
\begin{equation}
\begin{aligned}
    \dot{E} \equiv & \int \left[(E+P_\mathrm{tot}) v_r - B_r(\boldsymbol{B}\cdot\boldsymbol{v}) +\rho\Phi v_r\right] r^2 \dd \Omega,\\
    = & \int \Big[ \underbrace{\frac{1}{2}\rho v^2v_r}_\mathrm{Kinetic} + \underbrace{\rho h v_r}_\mathrm{Thermal} + \underbrace{B^2v_r-B_r(\boldsymbol{B}\cdot\boldsymbol{v})}_\mathrm{Magnetic}\\ & + \underbrace{\rho\Phi v_r}_\mathrm{Potential} \Big] r^2 \dd \Omega,\label{eq:edot}
\end{aligned}
\end{equation}
where $h\equiv (E_\mathrm{int}+P_\mathrm{gas})/\rho$ is the enthalpy per unit mass. The total energy flow is positive, indicating feedback, with hot gas dominating the energy budget and negligible contribution from cold gas (though not shown here). Magnetic, thermal, and kinetic energy successively dominate the energy flow from smaller to larger scales. Note that the potential energy flow, though positive, is the inflow of negative potential, and thus is not feedback. The amount of net energy flow increases with smaller accretor size $r_\mathrm{in}$, despite a lower accretion rate. Due to the computational cost, we can only trace the outflow to $\sim10^{3}\,r_\mathrm{in}$. Thus the energy outflow is essentially a superposition of outflows originating from different $r_\mathrm{in}$ ($2048,128,\text{and}\, 8\,r_\mathrm{g}$). The radial momentum flux is defined by
\begin{equation}
    \begin{aligned}
        \dot{p}\equiv & \int \rho v_r^2 r^2\dd\Omega,\\
        = & \underbrace{\int_{v_r<0} \rho v_r^2 r^2\dd\Omega}_{\dot{p}_\mathrm{in}} + \underbrace{\int_{v_r>0} \rho v_r^2 r^2\dd\Omega}_{\dot{p}_\mathrm{out}},\label{eq:momdot}
    \end{aligned}
\end{equation}
with both the inflow and outflow of the same sign. It is not the total flux in~\eq\ref{eq:mhd_momentum_eq} but indicates a possible constant that sets the scaling of the flow~\citep{Gruzinov2013arXiv1311.5813G}. The momentum flow is $\sim 3\times10^3\, M_\odot\,\mathrm{km\,s^{-1}\,yr^{-1}}$ (\figu\ref{fig:rad_flow}). The inflow is essentially independent of radius over a large dynamic range, especially within the Bondi radius ($\lesssim 10^5\,\mathrm{g}$), similar to previous hydrodynamic simulations. One important caveat is that the net mass and energy flow are not independent of radii because we cannot run long enough. The solutions here are \textit{not} statistical steady state at large radii. They may be at some intermediate radii but even that is not entirely clear.

\subsection{Time Evolution}

\figu\ref{fig:evo} shows the history of several key variables and their dependence on accretor size $r_\mathrm{in}$ of the fiducial MHD model along with a hydrodynamic case as a comparison. The magnetic field facilitates the inflow of the cold gas on scales $\sim \mathrm{kpc} - \mathrm{pc}$ via more efficient angular momentum transfer, leading to a higher mass accretion rate at $r=r_\mathrm{in}$ by a factor of $\sim 10$ and smaller angular momentum on all scales when compared with the hydrodynamic case. The accretion rate decreases with smaller accretor size $r_\mathrm{in}$, still roughly following the scaling of $\dot{M}\sim r^{1/2}$ with final accretion rate measured at $r_\mathrm{in}=8\,r_\mathrm{g}$ being $\sim 0.05\,M_\odot\,\mathrm{yr^{-1}}$. The direction of angular momentum varies on different scales, similar to the hydrodynamic case.

To determine the role of the magnetic field, we define normalized magnetic flux $\phi_\mathrm{B}$ by
\begin{equation}
    \phi_\mathrm{B}\equiv\frac{\sqrt{\pi}\int|B_r|r^2\dd\Omega}{r\sqrt{|\dot{M}|v_\mathrm{K}}},\label{eq:phi}
\end{equation}
and measure it at $r=r_\mathrm{in}$, similar to \citet{Ressler2020MNRAS.492.3272R} but different by a factor of $\sqrt{4\pi}$ because here magnetic permeability $\mu_\mathrm{m}=1$. In the Newtonian simulations, a rough threshold value to reach the MAD state is $\phi_\mathrm{B}\sim 2\pi\sqrt{v_\mathrm{K}/v_r}\big|_{r=r_\mathrm{in}}$, of the order of $6-10$ if $v_r\sim v_\mathrm{K}$ around the inner boundary. This is similar to the value we measured here, especially within $10^3\,r_\mathrm{g}$, where the accretion flow transitions to the hot phase. 

On the cold disk scale, there is a clear energetic outflow, unlike the hydrodynamic case. We measure the net energy flux at $r=128r_\mathrm{in}$, a scale $\gg r_\mathrm{in}$ so the energy feedback becomes independent of radius, and plot it in \figu\ref{fig:evo}. Despite decreasing mass accretion rate with smaller $r_\mathrm{in}$, the net energy flux becomes positive and increases with smaller accretor size, reaching $\sim 3\times10^{43}\,\mathrm{erg\,s^{-1}}$ when $r_\mathrm{in}=8\,r_\mathrm{g}$ (the last two columns in \figu\ref{fig:evo}). To quantify the strength of the feedback, we define a dimensionless feedback efficiency normalized using the Keplerian velocity at $r_\mathrm{in}$
\begin{equation}
    \eta\equiv\frac{\dot{E}(128r_\mathrm{in})}{\dot{M}v_\mathrm{K}^2(r_\mathrm{in})}\label{eq:eta},
\end{equation}
and plot it in \figu\ref{fig:evo}. The feedback efficiency is roughly a constant with $\eta\sim0.05-0.1$ when $r_\mathrm{in}\lesssim 10^3\,r_\mathrm{g}$. Therefore, though mass accretion rate scales down with smaller $r_\mathrm{in}$ following $\dot{M}\propto r_\mathrm{in}^{1/2}$, the energy outflow rate $\dot{E}\sim\eta\dot{M}v_\mathrm{K}^2\propto r_\mathrm{in}^{-1/2}$ and scales up. 

\subsection{Accretion Structure}\label{subsec:disk}

To better diagnose the structure and angular momentum transfer in the accretion disk, we define a cylindrical coordinate system ($R,\phi,z$) using the orientation of the cold disk and calculate the stresses. The total stress tensor can be decomposed into,
\begin{equation}
    \boldsymbol{\Pi}_\mathrm{tot}=\underbrace{\rho\boldsymbol{v}\boldsymbol{v}}_{\boldsymbol{\Pi}_\mathrm{kin}} + \underbrace{P_\mathrm{gas}\boldsymbol{\mathrm{I}}}_{\boldsymbol{\Pi}_\mathrm{therm}} + \underbrace{\left(\frac{\boldsymbol{B}\cdot \boldsymbol{B}}{2}\boldsymbol{\mathrm{I}}-\boldsymbol{B}\boldsymbol{B}\right)}_{\boldsymbol{\Pi}_\mathrm{mag}}.\\
\end{equation}
The kinetic stresses can be further separated into $\boldsymbol{\Pi}_\mathrm{kin}=\bar{\boldsymbol{\Pi}}_\mathrm{kin}+\delta\boldsymbol{\Pi}_\mathrm{kin}$ where the mean part $\bar{\boldsymbol{\Pi}}_\mathrm{kin}=\rho\langle\boldsymbol{v}\rangle\langle\boldsymbol{v}\rangle$ and the fluctuating part (the Reynolds stress) $\delta\boldsymbol{\Pi}_\mathrm{kin}=\rho\delta\boldsymbol{v}\delta\boldsymbol{v}$ with the fluctuating velocities $\delta \boldsymbol{v} = \boldsymbol{v} - \langle \boldsymbol{v}\rangle$. Similarly, the magnetic (Maxwell) stresses $\boldsymbol{\Pi}_\mathrm{mag}=\bar{\boldsymbol{\Pi}}_\mathrm{mag}+\delta\boldsymbol{\Pi}_\mathrm{mag}$ where $\bar{\boldsymbol{\Pi}}_\mathrm{mag}=|\langle\boldsymbol{B}\rangle|^2\boldsymbol{\mathrm{I}}/2-\langle\boldsymbol{B}\rangle\langle\boldsymbol{B}\rangle$ and $\delta\boldsymbol{\Pi}_\mathrm{mag}=|\delta\boldsymbol{B}|^2\boldsymbol{\mathrm{I}}/2-\delta\boldsymbol{B}\delta\boldsymbol{B}$ with fluctuating magnetic field $\delta \boldsymbol{B} = \boldsymbol{B} - \langle \boldsymbol{B}\rangle$. \figu\ref{fig:disk_radial} plots the radial profiles of mean and fluctuating stresses, velocities, and Alfv\'en velocities in the midplane ($|z/R|<0.2$). 

In the chaotic cold accretion region ($\gtrsim 30\,\mathrm{pc}$), apart from the frequent collisions between the filaments which also happen in the hydrodynamic case, the Maxwell stresses drive the angular momentum transfer efficiently with $\delta\Pi_{R\phi}^\mathrm{mag}\sim 0.1\rho v_\mathrm{K}^2$. The cold filaments lose angular momentum more quickly and thus enhance the accretion rate significantly, similar to previous works~\citep{Wang2020MNRAS.493.4065W}. Note that in the long run ($\gtrsim 100\,\mathrm{Myr}-1\,\mathrm{Gyr}$), though the dispersion of angular momentum is large, the mean angular momentum of all the gas is very close to zero due to the simulation setup. Thus the cold gas will finally fall into the sink region when feedback is absent. However, the mean angular momentum is not negligible at nearly any specific time point. So the effect of the magnetic field is to accelerate the process of angular momentum loss.

Though the accretion is chaotic on scales $\gtrsim 30\,\mathrm{pc}$, pure collisions and magnetic stresses cannot always transfer the angular momentum efficiently enough to keep the accretion chaotic. Instead, the stresses become less efficient in further transferring angular momentum within a few orbits inside $10^5\,r_\mathrm{g}$ ($30\,\mathrm{pc}$) so the cold flow is no longer chaotic but circularizes to a Keplerian disk in a radius $r_\mathrm{circ}$ $\sim 10$ times smaller than the hydrodynamic case. The exact $r_\mathrm{circ}$ depends on the exact time when we zoom in the simulations. The cold disk has large asymmetries, with disk orientation varying significantly on different scales due to the random distribution of the angular momentum of the infalling cold filaments, similar to previous hydrodynamic simulations.

To better understand the cold disk on scales $\sim 0.3-30$ pc ($10^3-10^5\,r_\mathrm{g}$), we plot the vertical profile of various quantities of the disk at $R\sim10^4\,r_\mathrm{g}$ in \figu\ref{fig:disk_vertical}. Similar to~\citet{Hopkins2024OJAp....7E..19H}, the disk midplane is cold with $T\sim 10^{-4} T_\mathrm{virial}$, dense with density $\sim 100$ times higher than the background, Keplerian with velocities $v_\phi\approx v_\mathrm{K}$, $v_R\sim-0.01v_\mathrm{K}$, $v_z\ll v_\mathrm{K}$, toroidal-field dominated with Alfv\'en velocities $v_{A,\phi}\sim 0.5v_\mathrm{K}$, $v_{A,R}\sim-0.02v_\mathrm{K}$, $v_{A,z}\ll v_\mathrm{K}$, and turbulent with $\delta v\sim\delta v_{A}\sim 0.5 v_{A} \sim 0.2v_\mathrm{K}$. Though not shown in the plot, the three components of the turbulent velocities and turbulent Alfv\'en velocities are similar in orders of magnitude with weak anisotropy. Strong Maxwell and Reynolds stress efficiently transfer the angular momentum with the dominant term in the midplane being $\bar{\Pi}_{R\phi}^\mathrm{mag}\sim 0.01 \rho v_\mathrm{K}^2>0$, indicating angular momentum loss. Above and below the disk surface, the vertical and radial velocities are outward with considerable vertical magnetic flux. There are strong wind stresses, i.e., the $z\phi$-component of the Maxwell stress, with $\delta\Pi_{z\phi}^\mathrm{mag} \sim \bar{\Pi}_{z\phi}^\mathrm{mag} \sim 0.01 \rho v_\mathrm{K}^2$ for $|z/R|\gtrsim0.5$. Though the mean part and the fluctuating part typically have opposite signs, the total stress is still quite strong. Therefore, the strong outflow/wind efficiently removes angular momentum from the disk. This is different from~\citet{Hopkins2024OJAp....7E..19H}, where the vertical velocity is primarily (weak) inflow.

Similar to \citet{Hopkins2024OJAp....7E..19H}, the strong magnetic field is due to flux freezing and advection. However, the strong toroidal field can be expelled by Parker-like buoyancy-related instabilities very quickly. We note the dynamics may not be correctly captured here since we do not resolve the thermal scale height of the cold disk. For the fluctuations of the magnetic field, we are not in the regime of the ``traditional'' MRI since the magnetic field is too strong. However, the disk is still unstable to other types of MRI~\citep{Balbus1992ApJ...400..610B, Pessah2005ApJ...628..879P}.

In the turbulent hot accretion flow region ($\lesssim10^3\,r_\mathrm{g}$), the accretion is less disk-like and highly turbulent with $\delta v\sim v_\phi \sim 0.5 v_\mathrm{K}$ and $v_R\sim -0.1v_\mathrm{K}$ in the midplane. The magnetic field is also entangled with $\delta v_{A}\sim 0.5 v_\mathrm{K}$ but has a coherent mean vertical magnetic field crossing the midplane with $v_{A,z}\sim 0.1 v_\mathrm{K}$. There are still strong Maxwell and Reynolds stresses ($\delta\Pi_{R\phi}^\mathrm{mag} \sim \delta\Pi_{R\phi}^\mathrm{kin} \sim 0.01 \rho v_\mathrm{K}^2$) driving the hot gas accretion and hot outflow. \figu\ref{fig:theta_flux} plots the angular profile of mass, energy, and momentum flux on scale $\sim 0.3\,\mathrm{pc}(10^{3}\,r_\mathrm{g})$. The inflows are primarily in the midplane ($\sim 60^\circ-120^\circ$) while the outflows are in the polar region ($\lesssim 30^\circ$ and $\gtrsim 150^\circ$). There are both strong mass inflow and outflow, leading to a smaller net flow. The energy outflow in the polar region is much larger than the energy inflow in the midplane. The momentum flux is similar in orders of magnitude at all angles.

\subsection{Properties of Outflow}\label{subsec:outflow}

We here analyze the strong outflow emerging in our simulations (\figus\ref{fig:zoom}, \ref{fig:logr}, and \ref{fig:theta_flux}) in more detail. Following previous works \citep[e.g.,][]{Blandford1982MNRAS.199..883B, Zhu2018ApJ...857...34Z, Zhu2024MNRAS.528.2883Z}, we take a look at the four conserved quantities in a steady axisymmetric MHD flow. First, the induction equation $\nabla\times(\boldsymbol{v}\times\boldsymbol{B})=0$ implies that the poloidal components $\boldsymbol{B}_\mathrm{\bold{p}}=\boldsymbol{B}_r+\boldsymbol{B}_\theta=\boldsymbol{B}_R+\boldsymbol{B}_z$ and $\boldsymbol{v}_\mathrm{\bold{p}}=\boldsymbol{v}_r+\boldsymbol{v}_\theta=\boldsymbol{v}_R+\boldsymbol{v}_z$ are in the same direction. The velocity thus can be related to the magnetic field by
\begin{equation}
    \boldsymbol{v}=\frac{k\boldsymbol{B}}{\rho}+\boldsymbol{\omega}\times\boldsymbol{R},
\end{equation}
where
\begin{equation}
    k=\frac{\rho\boldsymbol{v}_\mathrm{\bold{p}}}{\boldsymbol{B}_\mathrm{\bold{p}}},
    \label{eq:k}
\end{equation}
the mass loading parameter, and
\begin{equation}
    \omega=\frac{v_\phi}{R}-\frac{kB_\phi}{\rho R},
    \label{eq:omega}
\end{equation}
the angular velocity, are two conserved quantities along a streamline. From the angular momentum and total energy equations, we have the third constant, the specific angular momentum, 
\begin{equation}
    l=R(v_\phi-\frac{B_\phi}{k}),
    \label{eq:l}
\end{equation}
and the fourth constant, the Bernoulli parameter,
\begin{align}
    \label{eq:Be}
    \mathcal{B}&\equiv \frac{E+P_\mathrm{tot}}{\rho}-\frac{(\boldsymbol{B}\cdot\boldsymbol{v})(\boldsymbol{B}_\mathrm{\bold{p}}\cdot\boldsymbol{v}_\mathrm{\bold{p}})}{\rho|\boldsymbol{v}_\mathrm{\bold{p}}|^2}+\Phi,\\
    &=\frac{1}{2}v^2 + h + \frac{B_\phi^2}{\rho}-\frac{B_\phi v_\phi}{k}+\Phi.
\end{align}

We plot these variables together with a variety of velocities along a series of streamlines in the time and azimuthally-averaged outflow in \figu\ref{fig:disk_wind}. The outflow is super-Keplerian, supersonic, and super-Alfv\'enic and concentrates toward the pole. The conserved quantities ($k,\omega,l,\mathcal{B}$) are roughly constant within $\sim 50\%$ along the streamlines. This suggests that the outflow here is similar to the traditional steady-wind solution, though the accretion disk is highly turbulent and the vertical magnetic flux is weak. We note that winds are launching from a large range of radii due to our zoom-in strategy. Here we only pick the wind with launching point $r\sim 10^3\,r_\mathrm{g}$ and trace the streamline to the position where there is causality between the wind origin and the wind front.

\subsection{Thermal Transition from Cold to Hot Gas}

As noted before, the cold gas is truncated at $\sim 10^3\,r_\mathrm{g}$, and the cold accretion flow transitions to hot gas. To diagnose this in more detail, we restart the simulation with $r_\mathrm{in}=128\,r_\mathrm{g}$ and add a passive scalar to trace the evolution of the cold gas via 
\begin{equation}
    \frac{\partial (\rho C)}{\partial t}+\nabla \cdot(\rho \boldsymbol{v}C) = 0,
    \label{eq:scalar}
\end{equation}
where $C$ is the specific density of the scalar species, which is set to $1$ for the cold gas and $0$ otherwise. \figu\ref{fig:scalar} plots a slice of the accretion flow after an evolution of 10 kyr, corresponding to $\sim50$ orbits at $10^3\,r_\mathrm{g}$, and a hydrodynamic case for reference. In contrast to the hydrodynamic run, a considerable fraction of cold gas is converted to hot gas in the MHD run. We also run a test simulation with 14 levels of mesh refinement and $r_\mathrm{in}=512\,r_\mathrm{g}$ restarting directly from the largest-scale run (the run with 8 levels of mesh refinement and $r_\mathrm{in}=2^{15}\,r_\mathrm{g}\approx 10\,\mathrm{pc}$) and evolve it for 100 kyr. Though not shown here, there is a similar (though slightly smaller) truncation radius around $\sim1-2\times10^3\,r_\mathrm{g}$. 

This thermal transition is striking but still physically possible. As we have shown above, the magnetic field is toroidal field-dominated in the cold disk but poloidal field-dominated in the hot accretion flow. The cold gas may be magnetically arrested~\citep{Narayan2003PASJ...55L..69N, Tchekhovskoy2011MNRAS.418L..79T} or disrupted~\citep{Zhu2024MNRAS.528.2883Z}. As we see in \figu\ref{fig:logr}, the cold disk starts fragmenting into small pieces within $\sim 10^4\,r_\mathrm{g}$. In addition, the cooling time of the cold gas is \textit{not} significantly short due to a lower density supported by the magnetic field, but increases to $t_\mathrm{cool}\Omega_\mathrm{K}\sim 0.1$, especially on smaller scales (see \figu\ref{fig:rad}). There is a sharp increase in cooling time at $10^3\,r_\mathrm{g}$, which may be related to the transition from cold gas to hot gas. Furthermore, though not shown here, there is always a considerable fraction of intermediate gas with an even longer cooling time.

We note that there could be a cold disk within $10^3\,\mathrm{r_g}$ if there was a direct cold cloud/filament infall with incredibly small angular momentum ($r_\mathrm{circ}\ll 10^3\,\mathrm{r_g}$). By restarting a zoom-in simulation at a time point when there is direct gas infall, we find a small disk forms in the center. However, this situation is very rare in our simulations. Most of the gas inflow has a larger $r_\mathrm{circ}$, forming a disk and further accreting via angular momentum transfer due to Maxwell and Reynolds stress in the disk. Therefore another way to view the result here is, given a cold disk found in our simulations and a MAD-like hot gas accretion flow in the center as initial conditions, the cold phase of the disk cannot extend to a smaller scale for at least hundreds of orbits, despite efficient angular momentum transfer.

\subsection{Dependence on Initial Conditions and Resolutions}

The large-scale initial conditions may affect the accretion rate. We here investigate the effects of initial conditions by varying a series of parameters including:
\begin{itemize}
    \item strength of the initial large-scale magnetic field: plasma $\beta=10^2$, $10^4$ and hydrodynamic case.
    \item range of wavelength of the initial density perturbation and entangled magnetic field: $\lambda = [20,10],[10,5],\text{and}\,[5,3]\,\mathrm{kpc}$ from large to small scale.
\end{itemize}
These runs are performed with a resolution of $128^3$ cells per level (half that of the fiducial run) and a temperature floor of $T_\mathrm{floor}=2\times10^{5}\,\mathrm{K}$. We use 8, 12, 16, 20, and 24 levels of mesh refinement. We also convert the Newtonian MHD and hydrodynamic runs to GR runs with a fixed black hole spin $a=0.9375$ for the final level of runs. Detailed analysis of the GRMHD runs will be carried out in future work.

\figu\ref{fig:proj_beta} shows the projected density for the runs with half of the resolution of the fiducial run, $\lambda=20-10\,\mathrm{kpc}$, and initial $\beta=10^4$ and $\beta=10^2$. \figu\ref{fig:rad_beta} shows the radial profiles of $\beta$ for the two cases. When initial $\beta=10^4$, the effect of the magnetic field is quite weak, with the accretion structure similar to the hydrodynamic case. There are fewer filaments and a larger disk than in the case with initial $\beta =10^2$. The reason is that the accretion flow on scales $\gtrsim 10^5\,r_\mathrm{g}$ is still weakly magnetized. Though not shown here, the cold disk on the scales $\sim 10\,\mathrm{pc}$ is still magnetically supported with scale height $H\sim r$ and $\beta\sim 10^{-2}\ll 1$. The magnetic field in the hot gas is also finally saturated to $\beta\sim1$ for the cases with initial $\beta=10^4$ when $r\lesssim 10^4\,\mathrm{g}$ and there is still strong feedback with the feedback efficiency still being $\sim 0.05-0.1$.

The mass accretion rates for these various runs are summarized in \figu\ref{fig:scaling}. The accretion rates are insensitive to large-scale turbulence. When plasma $\beta\sim 10^4$, the effects of magnetic fields tend to be weaker, leading to a mass accretion rate similar to the hydrodynamic case though the cold gas is still highly magnetized. A stronger large-scale initial magnetic field tends to increase the mass accretion rate. Note that the accretion rate varies considerably during the evolution of each system. Here we only pick one snapshot of the largest-scale simulations and run zoom-in simulations as a representative case for each parameter combination. We do not find a tight correlation between the accretion rate and the parameters $\beta$ and $\lambda$. Instead, the final accretion rates are more sensitive to the instantaneous accretion rate on $\sim 10^5\,r_\mathrm{g}$. The scaling of $\dot{M}\propto r^{1/2}$ still roughly holds from $10^5\,r_\mathrm{g}$ to $10\,r_\mathrm{g}$, regardless of the strength of the magnetic field and structure of the accretion flow. However, general relativistic effects can break the scaling and considerably change the accretion rate, which we will investigate in more detail in future work. 

\section{Discussion}\label{sec:discussion}

The simulations presented here reveal various accretion modes on different scales:
\begin{itemize}
    \item Chaotic cold accretion along magnetized filaments on scales $\sim 0.03-3$ kpc ($10^5-10^7\,r_\mathrm{g}$)
    \item Highly magnetized cold disk accretion on scales $\sim 0.3-30$ pc ($10^3-10^5\,r_\mathrm{g}$)
    \item Magnetized turbulent hot gas accretion on scales $\lesssim 0.3\,\mathrm{pc}$ ($10^3\,r_\mathrm{g}$)
\end{itemize}
Below we discuss these different accretion modes and their implications.

\subsection{Filamentary Accretion}

The filamentary accretion on scales of hundreds of parsec is a structure distinct from the hydrodynamic simulations. The accretion flow is similar to the chaotic cold accretion~\citep{Li&Bryan2012ApJ...747...26L, Gaspari2013MNRAS.432.3401G} but more filamentary due to the presence of the magnetic field. The cold filaments are magnetized with $\beta\sim 1$ and $B\sim 10^2\,\mathrm{\mu G}$ (\figus\ref{fig:rad} and \ref{fig:radv}). The magnetic field mostly aligns with the major axis of the filaments but is sometimes bent significantly (\figu\ref{fig:logr}). The filaments we find are broadly similar to those found in recent work by~\citet{Wang2020MNRAS.493.4065W} and \citet{Fournier2024arXiv240605044F}, who focused on the formation and evolution of the cold filaments. On this scale, angular momentum transport is produced both by frequent collisions between the filaments (which also happens in the hydrodynamic case), and large-scale Maxwell stresses (\figu\ref{fig:disk_radial}). The cold filaments lose angular momentum more quickly and thus enhance the accretion rate significantly. Though the filamentary accretion is prominent on larger scales, a disk or torus still dominates the accretion on the parsec scale. Note that in the long run ($\gtrsim 100\,\mathrm{Myr}-1\,\mathrm{Gyr}$), though the dispersion of angular momentum is large, the mean angular momentum of all the gas is very close to zero due to the simulation setup. In the future, it will be interesting to investigate the accretion structure when there is net total angular momentum, which may be more realistic. 

\subsection{Formation of Highly Magnetized Cold Disk}

Though cold gas accretion is overall chaotic and turbulent at large radii, the cold filaments do not fall directly onto the SMBH but form a cold, thick, highly-magnetized, toroidal field-dominated disk or torus in the inner tens of parsec (or, in the most extreme cases, in the inner few pc) (\figus \ref{fig:zoom}, \ref{fig:logr}, and \ref{fig:proj_beta}). Unlike the hydrodynamic case, accretion continues due to strong Reynolds and Maxwell stresses in the cold disk. This disk is similar to the ``magnetically levitating'' accretion disks around SMBHs reported by \citet{Gaburov2012ApJ...758..103G} (see also earlier shearing box work on disks with strong toroidal fields by \citealt{Johansen2008}). Similar disks are also reported in \citet{Hopkins2024OJAp....7E..18H, Hopkins2024OJAp....7E..19H}, where they followed the accretion onto a SMBH in cosmological simulations from $\sim\,\mathrm{Mpc}$ scales at a time when it accretes as a bright quasar to $< 100\,\mathrm{au}\sim 600\,r_\mathrm{g}$ and conducted a detailed analysis of the accretion flow. Though the accretion rate relative to the Eddington limit is very different from the situation here, they also found a highly magnetized, toroidal field-dominated cold disk with the driver of accretion being the strong magnetic stress. They do not, however, find the strong outflows or the transition from cold gas to hot gas found here.

Can the cold disk be so highly magnetized? Parker-like buoyancy instabilities in principle can expel the toroidal flux very efficiently and keep $\beta\sim1$~\citep{Salvesen2016MNRAS.460.3488S}. Our current MHD simulations do not resolve the thermal scale height of the cold disk and thus may not correctly capture those instabilities. It is important to stress that this could change the ability of the cold gas to transition to a hot flow via magnetic heating at small radii as we found here and discuss in the next section (\figu~\ref{fig:logr}). The reason is that the efficiency of such a transition is sensitive to the cooling rate of the cold gas and thus its density and vertical structure/support.

\subsection{Magnetized Hot Gas Accretion}

Hot gas dominates the accretion within $\sim 10^3\,r_\mathrm{g}$. The accretion bears many similarities to the magnetized hot gas accretion in~\citet{Ressler2020MNRAS.492.3272R}. The gas is essentially in a MAD state with coherent magnetic flux. The basic flow structure is inflow from the midplane and outflow to the polar region. The hot gas has density $\rho\propto r^{-1}$ (\figu\ref{fig:rad}), similar to the turbulent accretion flow identified by constant momentum flux with $\rho\propto r^{-1}$ and $\dot{M}\propto r^{1/2}$~\citep{Gruzinov2013arXiv1311.5813G}, instead of the spherically symmetric Bondi flow with $\rho\propto r^{-3/2}$ or the Convection Dominated Accretion Flow (CDAF) with $\rho\propto r^{-1/2}$~\citep{Quataert2000ApJ...539..809Q}. There is some evidence that this scaling is universal if the accretion is nearly adiabatic and modestly turbulent, regardless of magnetization, as is reported in various contexts and with varying physics~\citep[e.g.,][]{Pang2011MNRAS.415.1228P, Yuan2012ApJ...761..129Y, Yuan2012ApJ...761..130Y, Yuan2015ApJ...804..101Y, Ressler2018MNRAS.478.3544R, Xu2019MNRAS.488.5162X, White2020ApJ...891...63W, Ressler2020ApJ...896L...6R, Guo2020ApJ...901...39G, Yang2021ApJ...914..131Y, Lalakos2022ApJ...936L...5L, Xu2023ApJ...954..180X}.

Previous hydrodynamic simulations with multiphase gas, heating, and cooling produce a similar accretion rate $\propto r_\mathrm{in}^{1/2}$~\citep{Guo2023ApJ...946...26G}. However, the hot gas dynamics are very different in those simulations without magnetic fields: there are no strong outflows of hot gas when cooling produces a significant cold disk. Instead, the hot gas either directly flows into smaller radii or settles relatively smoothly onto the cold disk and the mixing of hot and cool gas depletes the amount of hot gas at smaller radii. However, in the MHD simulations here, the accretion is turbulent, with mean inflow in the mid-plane and outflow in the polar region even for the case with initial large-scale $\beta\sim 10^4$. The hot gas dynamics are more similar to the traditional picture in which the reduction in the accretion rate relative to the Bondi rate ($\dot{M} \propto r^p$ for some $1 > p > 0$) is because of outflows suppressing the accretion rate at small radii \citep{Blandford1999MNRAS.303L...1B}, with some other conserved quantity setting the density profile and thus $\dot{M}(r)$ \citep[e.g.,][]{Narayan2000ApJ...539..798N, Quataert2000ApJ...539..809Q, Gruzinov2013arXiv1311.5813G}.

\subsection{Contribution of Hot Gas and Cold Gas}

Accretion of cold gas is the key in current high-resolution simulations of how active galactic nucleus (AGN) feedback balances cooling in clusters ~\citep[][etc]{Li&Bryan2012ApJ...747...26L, Gaspari2013MNRAS.432.3401G}. These models are essentially a limit cycle in which cold gas accretion generates feedback that suppresses cooling for a period of time (a few cooling times of hot gas), and then a new burst of cold gas accretion sets in. We find, by contrast, that while a cold disk with efficient accretion is present on parsec scales, it does not generally contribute directly to the accretion rate at small radii, which is dominated by the hot gas. Using passive scalars, we verified that the hot gas is produced directly from the cold gas (\figu\ref{fig:scalar}). A possible explanation is that the accumulation of magnetic flux leads to the formation of a strongly magnetized disk and ultimately to the transition of the cold gas to a hot flow. Magnetic fields can facilitate this in two ways: by increasing the scale height of the cold gas and thus decreasing its density and cooling rate, and by providing strong magnetic heating (via reconnection) of the cold gas to promote thermal runaway to a hot phase. It is important to stress, however, that the cold gas dynamics is not as well captured in our simulations as that of the hot gas. For example, we are not properly resolving the mixing layer between the cold and the hot phase, so the cold gas growth or destruction is not well captured \citep[e.g.,][]{Fielding2020ApJ...894L..24F, Gronke2022MNRAS.511..859G} 

\subsection{Implications for Subgrid Model}

One of the major goals of this work is to provide a better subgrid model of mass accretion rate on the horizon scale based on the properties on the Bondi scale. Using a series of hydrodynamic simulations, \citet{Guo2023ApJ...946...26G} demonstrated that the scaling $\dot M \sim \dot{M}_\mathrm{B} (r_\mathrm{g}/r_\mathrm{B})^{1/2}$ may be a good prescription for the accretion rate on event horizon scales when the accretion is hot gas dominated. The accretion rate is a hundred times smaller than the Bondi accretion rate prediction. It is remarkably consistent with $\dot{M}\sim(3-20)\times10^\mathrm{-4}\,M_\odot\,\mathrm{yr^{-1}}$ from EHT observations~\citep{M87EHT_VIII_2021ApJ...910L..13E}.

In the MHD simulations presented here, the radial dependence of mass accretion rate is still well described by the $r_\mathrm{in}^{1/2}$ scaling, especially when cold inflow is negligible (e.g., $r\lesssim 10^{3}\,r_\mathrm{g}$), similar to \citet{Ressler2020MNRAS.492.3272R}. However, unlike~\citet{Ressler2020MNRAS.492.3272R}, where the effect of the magnetic field on accretion rate is surprisingly small, the mass accretion rate in the MHD simulations presented here is systematically higher than the hydrodynamic simulations by a factor of $\sim 10$ for initial $\beta \sim 10^2$, as summarized in \figu\ref{fig:scaling}. At higher initial $\beta$ the accretion rate is more similar to the hydrodynamic simulations. The increase in accretion rate for stronger initial fields is because the magnetic field increases angular momentum transfer in the cold filamentary gas. The final accretion rate is typically $\sim 10^{-2}\,M_\odot\,\mathrm{yr^{-1}}$ in our preliminary GR simulations that go all the way to the horizon (\figu\ref{fig:scaling}). This accretion rate is somewhat higher than the current estimates of M87* but still far below the Eddington limit with $\dot{M}\sim 10^{-4}\,\dot{M}_\mathrm{Edd}$ where $\dot{M}_\mathrm{Edd}\approx130\, M_\odot\,\mathrm{yr^{-1}}$ is the Eddington accretion rate for M87*. Given the chaotic nature of the flow and the lack of full feedback between small and large scales in our simulations, the accretion rate realized in our simulations should not necessarily be exactly that realized in M87 today; given the near-impossibility of correctly predicting the instantaneous accretion rate ``today'', it is encouraging that the value we find is of the same order of magnitude as that inferred observationally.

The vacuum sink boundary conditions in our simulations mean the absence of radial pressure support, leading to a negligible outflow rate through the inner boundary and increasing the resulting accretion rate at small radii. This boundary condition is reasonable only if $r_\mathrm{in}$ is the event horizon of a non-spinning black hole. We carried out a series of GRMHD simulations with a fixed spin $a=0.9375$ to investigate the effect of spin. The accretion rate through the horizon is also shown in \figu\ref{fig:scaling}. The final accretion rate is reduced by a factor of a few if there is a spinning black hole. We will perform a detailed analysis of the GRMHD simulations and investigate the dependence on the spin in future studies.

In the parameter space we explored, we suggest a subgrid model for the accretion rate,
\begin{equation}
    \dot{M}_\mathrm{acc}=\left(\frac{10 r_\mathrm{g}}{r_\mathrm{B}}\right)^{1/2}\dot{M}_\mathrm{B}.
\end{equation}
As Figure \ref{fig:scaling} shows, the exact normalization of this expression varies by a factor of $\sim 10$ depending on the initial magnetic strength in the intracluster medium at large radii.

\subsection{Feedback from Small Scales}

One important feature in the MHD simulations is the energetic hot supersonic outflow concentrated towards the poles which carries considerable feedback energy; such an outflow was not present in our analogous hydrodynamic simulations. Although the mass accretion rate decreases as we decrease the accretor size, the energy feedback increases, reaching $\dot{E}\sim 3\times10^{43}\,\mathrm{erg\,s^{-1}}$ when $r_\mathrm{in}=8\,r_\mathrm{g}$ (\figus\ref{fig:rad_flow} and \ref{fig:evo}). This corresponds to a roughly constant energy feedback efficiency of $\eta\sim0.05-0.1$ independent of $r_\mathrm{in}$. The feedback is concentrated towards the poles with a half-opening angle $\lesssim 30^\circ$ (\figu\ref{fig:theta_flux}). A key question is whether the energy feedback is enough to shut off the cooling flow. \citet{Luan2018ApJ...862...73L} presented a deep Chandra observation of M87 and estimated a total X-ray luminosity of $\sim3\times 10^{42}\,\mathrm{erg\,s^{-1}}$ within 10 kpc and $\sim 10^{43}\,\mathrm{erg\,s^{-1}}$ within 50 kpc excluding point sources (private communication). Thus the feedback we find is energetically sufficient to shut off the large-scale cooling flow in M87. Our current simulations cannot assess how well the feedback in fact couples to large radii; we will study this in more detail in future work. 

There is also a radial momentum feedback of $\sim 3\times10^3\, M_\odot\,\mathrm{km\,s^{-1}\,yr^{-1}}$, essentially independent of radius (\figu\ref{fig:rad_flow}). It corresponds to a speed of $\sim 10^4\,\mathrm{km\,s^{-1}}$, similar to typical assumptions in jet feedback models~\citep{Su2023MNRAS.520.4258S}. The momentum feedback is significant in the sense that it can drive a mass flux of $\sim 5\, M_\odot\,\mathrm{yr^{-1}}$ at the escape velocity of the galaxy ($\sim 600\,\mathrm{km\,s^{-1}}$). It is comparable to or higher than the momentum flux of $\sim5-20\times10^2\, M_\odot\,\mathrm{km\,s^{-1}\,yr^{-1}}$ in central starburst-driven outflows with star formation rate of $5-20\,M_\odot\,\mathrm{yr^{-1}}$ \citep{Schneider2020ApJ...895...43S} and much higher than $< 100 M_\odot\,\mathrm{km\,s^{-1}\,yr^{-1}}$ seen in outflows from star-forming disks with star formation rate surface density up to $<1 M_\odot\,\mathrm{kpc^{-2}\,yr^{-1}}$ \citep{Kim2020ApJ...900...61K}. 

\section{Summary}\label{sec:summary}

We have presented a series of MHD simulations of the fueling of SMBHs in elliptical galaxies, taking M87 as an example. Our results illustrate the interplay between heating, cooling, and accretion on various scales in massive galaxies and galaxy clusters. Our simulations successively zoom-in to the galactic nucleus and thus represent a realization of the small-scale accretion flow and the outflows it produces for a given set of conditions at large radii. A limitation of this approach is that we do not run our simulations long enough to reach a true statistical steady state in which the large-scales adjust to the feedback produced by smaller scales. Our main conclusions are
\begin{enumerate}
    \item The mass accretion rate is increased significantly in MHD simulations with large-scale initial $\beta\lesssim10^2$ by a factor of $\sim 10$ compared with analogous hydrodynamic simulations. The scaling of $\dot{M} \sim r^{1/2}$ still roughly holds from $\sim 10\,\mathrm{pc}$ to $\sim 10^{-3}\,\mathrm{pc}$ ($\sim 10\, r_\mathrm{g}$) with the accretion rate through the event horizon being $\sim 10^{-2}\, M_\odot\,\mathrm{yr^{-1}}$ (\figu\ref{fig:scaling}). In the parameter space we explored, we suggest that an accretion rate suppressed relative to the Bondi rate by $\sim \left({10 r_\mathrm{g}}/{r_\mathrm{B}}\right)^{1/2}$ is a more physical and accurate subgrid model of SMBH fueling by hot gas than the standard Bondi accretion rate assumption typically used in large-scale cosmological simulations.
    \item The accretion flow on scales $\sim 0.03-3\,\mathrm{kpc}$ is filamentary, similar to a chaotic cold accretion (\figu\ref{fig:zoom}). The cold filaments are magnetized with near free-fall velocities. This is due to efficient angular momentum transfer by Maxwell stresses and frequent collisions between the filaments.
    \item A smaller disk typically forms within $\sim 30\,\mathrm{pc}$ though the accretion is less coherent on larger scales. The disk is cold ($T\sim 10^4\,\mathrm{K}$), Keplerian ($v_\phi\approx v_\mathrm{K}$), geometrically thick ($H/R\sim0.5$) due to magnetic pressure $(v_{A}\sim 0.5v_\mathrm{K})$, highly magnetized ($\beta\sim10^{-3}$) with a primarily toroidal field, and turbulent ($\delta v\sim0.3v_\mathrm{K}$) with strong spiral arms (\figus\ref{fig:zoom} and \ref{fig:disk_vertical}).
    \item The cold disk is truncated and transitions to a turbulent hot accretion flow interior to $\sim 0.3\,\mathrm{pc}$ ($10^3\,r_\mathrm{g}$) (\figu\ref{fig:logr}). We attribute this to strong magnetic heating via, e.g., reconnection and suppressed cooling due to the lower densities in magnetically supported disks (\figu\ref{fig:rad}). We are not confident, however, that the cold gas dynamics is fully resolved, so further work on the thermal and dynamical properties of such strongly magnetized disks is critical.
    \item The hot accretion flow at the smallest radii in our simulations is essentially in a MAD state, with near-virial temperatures, $\delta v \sim v_\phi\sim 0.5 v_\mathrm{K}$, and a turbulent magnetic field with $\beta\sim1$ and a significant component of coherent vertical magnetic flux with the normalized $\phi_\mathrm{B}\sim6-10$ (\figus\ref{fig:rad}, \ref{fig:radv}, and \ref{fig:evo}). It is likely that on horizon scales the large-scale magnetic flux found here will produce a powerful jet, as is observed in M87.
    \item There is a strong outflow towards the poles driven by the magnetic field from the hot accretion flow at small radii (\figus\ref{fig:zoom}, \ref{fig:logr}, and \ref{fig:disk_wind}). The outflow is hot, magnetized, supersonic, super-Alfv\'enic, and super-Keplerian with $v_r\gtrsim v_A \sim c_\mathrm{s} \gtrsim v_\mathrm{K}$ (\figu\ref{fig:disk_wind}); the outflow is concentrated towards the poles with a half-opening angle $\lesssim 30^\circ$ (\figu\ref{fig:theta_flux}). The outflow energy flux increases with smaller accretor size, reaching $\sim 3\times10^{43}\,\mathrm{erg\,s^{-1}}$ for $r_\mathrm{in}=8\,r_\mathrm{g}$ with a corresponding momentum flux $\sim 3\times10^3\, M_\odot\,\mathrm{km\,s^{-1}\,yr^{-1}}$ (\figus\ref{fig:rad_flow} and \ref{fig:evo}). This energy outflow rate is in principle sufficient to fully balance cooling in the hot halo of M87.
\end{enumerate}

Simulations using more realistic heating models and GRMHD simulations with these more realistic physical ingredients will be carried out in the near future.

\section*{Acknowledgment}
We thank Phil Hopkins and Jono Squire for useful conversations. We thank the anonymous referee for the helpful comments and suggestions. This work was supported by a grant from the Simons Foundation (888968, E.C. Ostriker, Princeton University PI) as part of the Learning the Universe Collaboration.
JS acknowledges support from the Eric and Wendy Schmidt Fund for Strategic Innovation. EQ was supported in part by a Simons Investigator grant from the Simons Foundation and NSF AST grant 2107872. CGK was supported in part by NASA ATP grant 80NSSC22K0717.
The authors are pleased to acknowledge that the work reported on in this paper was substantially performed using the Princeton Research Computing resources at Princeton University which is consortium of groups led by the Princeton Institute for Computational Science and Engineering (PICSciE) and Office of Information Technology's Research Computing.
This work used the Delta system at the National Center for Supercomputing Applications through allocation PHY230165 from the Advanced Cyberinfrastructure Coordination Ecosystem: Services \& Support (ACCESS) program, which is supported by National Science Foundation grants \#2138259, \#2138286, \#2138307, \#2137603, and \#2138296~\citep{Boerner10.1145}.

\vspace{5mm}
\software{\athena~\citep{Stone2020ApJS..249....4S}}

%\appendix

\end{CJK*}

\bibliography{main}{}
\bibliographystyle{aasjournal}


\end{document}